\documentclass[9pt,preprint,nocopyrightspace,numbers]{sigplanconf}
\pdfoutput=1
\usepackage{amsmath}
\usepackage{color}
\usepackage{paralist}
\usepackage{comment}
\usepackage[ruled,vlined,linesnumbered]{algorithm2e}
\makeatletter
\newcommand{\removelatexerror}{\let\@latex@error\@gobble}
\makeatother
\usepackage{url}                  
\usepackage{listings}
\usepackage{multirow}
\usepackage{times}
\usepackage{array}
\usepackage[shortlabels]{enumitem}
\usepackage{amssymb}
\usepackage{adjustbox}
\usepackage{mathtools}
\usepackage{booktabs}
\usepackage{fixltx2e}

\newcommand{\ra}[1]{\renewcommand{\arraystretch}{#1}}

\usepackage[backgroundcolor=yellow,textsize=scriptsize]{todonotes}

\newcolumntype{x}[1]{>{\raggedright\arraybackslash}p{#1}}
\newcolumntype{L}[1]{>{\raggedright\let\newline\\\arraybackslash\hspace{0pt}}m{#1}}
\newcolumntype{C}[1]{>{\centering\let\newline\\\arraybackslash\hspace{0pt}}m{#1}}
\newcolumntype{R}[1]{>{\raggedleft\let\newline\\\arraybackslash\hspace{0pt}}m{#1}}

\newcommand{\TODO}[1]{\todo[inline]{#1}}

\usepackage{tikz}
\usetikzlibrary{positioning, calc}
\tikzset{circle node/.style = {circle,inner sep=1pt,draw, fill=white},
        X node/.style = {fill=white, inner sep=1pt},
        dot node/.style = {circle, draw, inner sep=2pt, fill=green}
}
\usepackage{tikz-3dplot}

\usetikzlibrary{patterns}

\usepackage[colorlinks=true,allcolors=blue,draft=false,debug=true]{hyperref}   

\usepackage{ifthen}

\newcommand{\DONE}[1]{}
\newcommand{\COMMENT}[1]{}

\newcommand{\figref}[1]{Fig.~\ref{Fi:#1}}

\newcommand{\theref}[1]{Theorem~\ref{Th:#1}}
\newcommand{\tabref}[1]{Table~\ref{Ta:#1}}
\newcommand{\secref}[1]{Section~\ref{Se:#1}}

\newcommand{\ssecref}[1]{Sec.~\ref{Se:#1}}
\newcommand{\appref}[1]{Appendix~\ref{Se:#1}}

\newcommand{\algref}[1]{Algorithm~\ref{Alg:#1}}

\newcommand{\slnref}[1]{\ref{Ln:#1}}

\newcommand{\figlabel}[1]{\label{Fi:#1}}
\newcommand{\deflabel}[1]{\label{De:#1}}
\newcommand{\thelabel}[1]{\label{Th:#1}}
\newcommand{\tablabel}[1]{\label{Ta:#1}}
\newcommand{\seclabel}[1]{\label{Se:#1}}

\newcommand{\applabel}[1]{\label{Se:#1}}

\newcommand{\alglabel}[1]{\label{Alg:#1}}

\newcommand{\lnlabel}[1]{\label{Ln:#1}}

\newcommand{\ignore}[1]{}

\newcounter{programlinenumber}

\newboolean{TR}
\setboolean{TR}{false}
\ifthenelse{\boolean{TR}}{

\newcommand{\TrOnly}[1]{#1}
\newcommand{\SubOnly}[1]{}
\newcommand{\TrOnlyInFootnote}[1]{#1}
\newcommand{\TrOnlyInTable}[1]{#1}}
{

\newcommand{\TrOnly}[1]{}
\newcommand{\SubOnly}[1]{#1}
\newcommand{\TrOnlyInFootnote}[1]{}
\newcommand{\TrOnlyInTable}[1]{}}

\newcommand{\hiddentext}[1]{}

\newcommand{\para}[1]{\vspace{3pt}\noindent\textbf{\textit{#1}}}

\newcommand{\tool}{{\small \textsc{SyFi}}}

\newcommand{\rangep}[3]{(#2 \leq #1 \leq #3)}

\newcommand{\ltpred}[2]{#1 \! < \! #2}

\newcommand{\scode}[1]{{\small \texttt{#1}}}

\renewcommand{\implies}{\Rightarrow}

\renewcommand{\phi}{\varphi}

\definecolor{graybck}{gray}{0.9}

\newcommand{\name}{{\small \textsc{ELE}}}
\newcommand{\alg}{{\small \textsc{SPEX}}}

\begin{document}

\setlength{\pdfpageheight}{\paperheight}
\setlength{\pdfpagewidth}{\paperwidth}

\conferenceinfo{CONF 'yy}{Month d--d, 20yy, City, ST, Country}
\copyrightyear{20yy}
\copyrightdata{978-1-nnnn-nnnn-n/yy/mm}
\copyrightdoi{nnnnnnn.nnnnnnn}
\setlength{\floatsep}{10pt plus 1.0pt minus 1.0pt}
\setlength{\textfloatsep}{10pt plus 2.0pt minus 2.0pt}

\newtheorem{definition}{Definition}
\newtheorem{claim}{Claim}
\newtheorem{theorem}{Theorem}
\newtheorem{challenge}{Challenge}

\title{Optimal Learning of Specifications from Examples}

\authorinfo{Dana Drachsler-Cohen}{Technion}{Israel}
\authorinfo{Martin Vechev}{ETH Z\"urich}{Switzerland}
\authorinfo{Eran Yahav}{Technion}{Israel}

\maketitle

\begin{abstract}
A fundamental challenge in synthesis from examples is designing a learning algorithm that poses the minimal number of questions to an end user while guaranteeing that the target hypothesis is discovered. Such guarantees are practically important because they ensure that end users will not be overburdened with unnecessary questions. 

We present {\alg}---a learning algorithm that addresses the above challenge. {\alg} considers the hypothesis space of formulas over first-order predicates and learns the correct hypothesis by only asking the user simple membership queries for \emph{concrete examples}. Thus, {\alg} is directly applicable to any learning problem that fits its hypothesis space and uses membership queries.

{\alg} works by iteratively eliminating candidate hypotheses from the space until converging to the target hypothesis. The main idea is to use the implication order between hypotheses to guarantee that in each step the question presented to the user obtains maximal pruning of the space. This problem is particularly challenging when predicates are potentially correlated.

To show that {\alg} is practically useful, we expressed two rather different applications domains in its framework: learning programs for the domain of technical analysts (stock trading) and learning data structure specifications. The experimental results show that {\alg}'s optimality guarantee is effective: it drastically reduces the number of questions posed to the user while successfully learning the exact hypothesis.
\end{abstract} 

\section{Introduction}\seclabel{Intro}
Over the last few years, programming by example (PBE) techniques have proved useful in a variety of application domains (e.g.,~\cite{Polozov:2015,Commutatitivity:2015, Singh:2016,Raychev:2016,Frankle:2016,Feser:2015,Barowy:2015,Singh:VLDB12,Gulwani:CACM12,Gulwani:2011:ASP:1926385.1926423,FlashExtract:14,BitManipulation:2010}). The goal of PBE approaches is to synthesize a hypothesis (e.g., a program \cite{BitManipulation:2010} or a specification  \cite{Commutatitivity:2015}) desired by an end user from answers of questions the synthesizer poses to that user. Thus, a prime objective for any PBE approach is to reduce the burden placed on the user. This means that it is critical to reduce the number of questions the end user has to answer, ideally, to a minimum. While existing approaches have focused on PBE engines that learn hypotheses in interesting domains, there has been little work on guaranteeing that the target hypothesis can be discovered with a minimal number of user questions. In fact, existing PBE approaches (e.g.,~\cite{BitManipulation:2010}) may ask the user an exponential number of questions even when a linear number would have sufficed, limiting the practical benefits of PBE.

\paragraph{This Work}
We present \alg, a new approach which ensures that for a given hypothesis space, the PBE engine will find the target solution with a \emph{minimal} number of questions posed to the end user. To obtain this result we had to address two challenges: (i) define the hypothesis space and identify key properties on its shape; in turn, this allows our search procedure to detect hypotheses whose testing enables maximum pruning of the search space, and (ii) uncover the place in the search where involving the user is most beneficial and thus we are guaranteed their involvement is reduced to a minimum.

Concretely, \alg\ considers: (i) a hypothesis space defined by formulas over first-order predicates, and (ii) membership questions which are posed to the end user; we note that simple membership questions of various flavors are a staple of PBE approaches and are suitable for end users to answer. In our setting, membership questions are simple questions on concrete examples with the answer determining whether a predicate is relevant to the hypothesis we are trying to learn. Any synthesis problem which has the same hypothesis space and considers membership questions like \alg\ can immediately benefit from our results.

\paragraph{\alg\ Operation}
To learn a hypothesis, \alg\ maintains a strict formula $\varphi$ which logically implies the hypothesis we are trying to discover, and gradually attempts to relax it. At each step, \alg:
\begin{inparaenum}[(i)]
  \item considers a minimally relaxed hypothesis $\hat{\varphi}$,
  \item generates a distinguishing input $e$ for $\varphi$ and $\hat{\varphi}$, \label{steptwo}
  \item asks the user for $e$'s correct output (via a membership question), and \label{stepthree}
  \item accordingly decides whether $\varphi$ can be relaxed to $\hat{\varphi}$.
\end{inparaenum}
Unfortunately, this approach only works for independent predicates, since for dependent predicates step (\ref{steptwo}) can fail. In fact, an even more restricted version of this approach (one that considers special cases of independent predicates) was proposed by ~\cite{Angluin:1993}. They show that for other cases (e.g., dependent predicates), not only this approach fails but also that it cannot be accomplished with a polynomial number of questions.

This is where \alg's key technical novelty lies in: we show how to proceed at step (\ref{steptwo}) even in the case of dependent predicates while guaranteeing we ask a minimal number of questions at step (\ref{stepthree}). The beauty of this approach is that the number of questions is fully adaptable to the choice of predicates. That is, \alg's guarantees are not obtained by the general worst-case, but by the worst-case of the hypothesis space determined by the \emph{given} set of predicates. Such guarantees are also known as the teaching dimension of the hypothesis space~\cite{Goldman:1995}. In addition, we present another result which further characterizes hypothesis spaces where the number of questions presented by \alg\ is guaranteed to be linear (our result subsumes works that consider independent predicates).

In addition to minimality guarantees, we show our framework can accommodate interesting application domains: we expressed two different synthesis problems in \alg: one where we learn technical patterns (i.e., programs used in stock trading) and one where we learn data structure specifications. We also show experiments demonstrating that \alg\ significantly reduces the number of questions posed to the user, when compared to current approaches.

\paragraph{Main Contributions} The main contributions are:
\begin{itemize}
\item \alg: an interactive PBE system that learns a target hypothesis expressed by formulas over first-order predicates using a \emph{minimal} number of membership questions.
\item A result which states that for a certain useful class of hypothesis spaces, the number of examples presented to the user is \emph{linear}.
\item Instantiation of \alg\ on two application domains: technical analysis patterns and data structure specifications. We show that our guarantee is practically useful: \alg\ asks the user significantly fewer questions than current approaches.
\end{itemize}

%

\section{Overview}\seclabel{Overview}
In this section, we informally explain \alg. Formal details are provided in later sections.

\subsection{Exact Learning from Examples}\seclabel{overview1}
We address the problem of exact learning from examples (\name). In {\name}, a synthesizer (learner) tries to learn a concept by presenting examples for classification by a user (teacher). The user may also provide initial sets of positive and negative examples. Technically, given a domain of examples $D$, a concept $C \subseteq D$ is a subset of the example domain. An example $e \in C$ is referred to as a \emph{positive} example, and an example $e \in D \setminus C$ is a \emph{negative example}. For instance consider:
 \begin{itemize}
   \item A domain $D=\{(x,y)\mid~0\leq x \leq 4,0 \leq y \leq 4\}$ .
   \item A concept $C=\{(1,1),(1,2),(2,1),(2,2)\}$.
   \item A single initial example $(2,2)$, which is positive.
 \end{itemize}
 \figref{Simple} shows $D$ and $C$ (whose points are marked with bold points).

\para{The Challenge of Exact Learning}
Exact learning algorithms have to learn a single concept. However, the initial examples the user provides are often consistent with many concepts. In our example, there are many subsets of $D$ that contain the example $(2,2)$. To isolate the correct concept, exact learning algorithms are allowed to \emph{present questions to the user}. This enables pruning concepts inconsistent with the new examples until a single concept remains.

\para{Membership Queries} Two common kinds of questions presented by exact learning algorithms are \emph{membership queries} and \emph{validation queries}.
Membership queries present examples (elements from $D$) and ask whether they belong to the concept, while validation queries present concepts and ask whether these are the correct one. Unfortunately, in many domains, validation queries are complex, error-prone, or impossible for the user to understand. For such domains, it is desirable to limit the questions to membership queries only, i.e., limit the setting to \name: exact learning from examples.

\para{Predicate-defined Concepts}
Often, concepts are conveniently specified using their \emph{features} or properties. In this work, we assume that concepts are defined using arbitrary predicates.
For example, we can express the concepts of $D$ from our running example with the following set of predicates:
$$S=P_x \cup P_y \cup \{x=y\}$$ where $P_x$ consists of predicates capturing intervals of $x$: $P_x = \{\rangep{x}{0}{2}, \rangep{x}{1}{3}, \rangep{x}{2}{4}\}$
and $P_y$ is defined identically with respect to $y$. A concept satisfies the predicate $a \leq x \leq b$ if all its points $(x,y)$ satisfy that $x$ is between $a$ and $b$, and a concept satisfies the predicate $x=y$ if all its points take the form of $(x,x)$.
Using these predicates, the concept we wish to learn (the one depicted in \figref{Simple}) is expressible by the formula: $$\varphi_C=(1\leq x \leq 3) \land (0 \leq x \leq 2) \land (1\leq  y\leq 3) \land (0 \leq y \leq 2)$$


\para{Finding the Correct Concept} To learn the concept formula $\varphi_C$, one can present examples to the user and prune the inconsistent concepts until a single concept remains. Many classical exact learning algorithms may be used only if the predicates are independent (see \secref{Related}). Unfortunately, practical application domains often define specifications over abstract properties that may be dependent.

One approach~\cite{BitManipulation:2010} does address this challenge by iteratively picking two non-equivalent concepts, showing an input that distinguishes them, asking the user for the correct output, and pruning the inconsistent concepts accordingly. Unfortunately, since the two concepts are selected \emph{arbitrarily}, the number of concepts pruned after a single question may be small, which can result in \emph{presenting an exponential number of questions to the user even when a linear number would suffice} (see~\secref{eval}). In contrast, we leverage the partial-order between the concepts to pick two ``close'' concepts at each step. This guarantees that, overall, our approach always asks the user the minimal number of questions.

In the next two sections, we focus on learning concepts that can be expressed as conjunctions. We then show that learning disjunctions is dual and that learning conjunctions may be used to learn DNF formulas (and thus any specification).

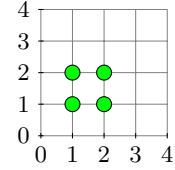
\begin{figure}
\centering
\begin{tikzpicture}
\def\sf{0.42} 
\draw[step=\sf *1cm,gray,very thin] (0,0) grid (\sf *4,\sf *4);
\foreach \x in {0,1,2,3,4}
    \draw (\sf * \x cm,1pt) -- (\sf * \x cm,-1pt) node[anchor=north] {$\x$};
\foreach \y in {0,1,2,3,4}
    \draw (1pt,\sf *\y cm) -- (-1pt,\sf *\y cm) node[anchor=east] {$\y$};	
 \node [dot node] at (\sf *1,\sf *1) {};
 \node [dot node] at (\sf *1,\sf *2) {};
 \node [dot node] at (\sf *2,\sf *1) {};
 \node [dot node] at (\sf *2,\sf *2) {};
 \end{tikzpicture}
\caption{Concept to be learned is $x \in [1,2] \land y \in [1,2]$.}\figlabel{Simple}
\end{figure}

\begin{figure*}
\centering
  \includegraphics[page=2,height=5cm,width=17cm,clip=true,trim=0pt 20pt 0pt 0pt]{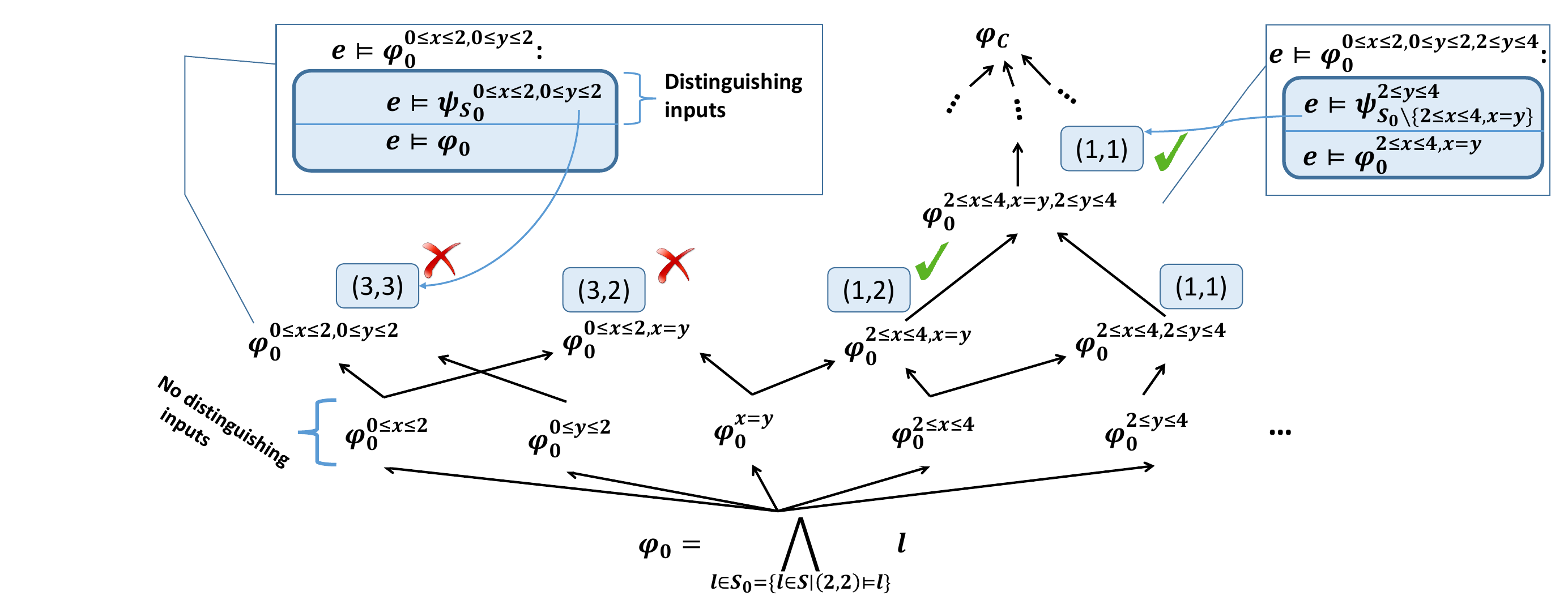}
  \caption{The partially-ordered concept space that \alg\ traverses along and distinguishing inputs showed by \alg.}
  \figlabel{conspace}
\end{figure*}

\subsection{Our Approach: a Guided Traversal in the Concept Space}\seclabel{congraph}
Intuitively, we follow the classic techniques as described in \secref{Intro}, which use the structure of the formula to efficiently check if a candidate hypothesis can be relaxed by asking the user about a distinguishing input, namely an input whose output changes when relaxing the formula. However, in the general setting of predicate-defined concepts, some hypotheses may not have such distinguishing input. 
The main idea of our approach is to leverage a partial order between concepts to find a minimal number of inputs that, together, act as the nonexistent distinguishing input.

Conceptually, \alg\ traverses along the \emph{partially-ordered} space of concepts consistent with the examples to find the correct concept formula $\varphi_C$. The partial-order is defined as follows: two formulas $\varphi,\psi$ satisfy $\varphi \leq \psi$ if $\varphi$ contains all of $\psi$'s predicates (and possibly additional predicates). This order induces a graph of the concepts consistent with the observed examples: the nodes are the consistent concepts, captured by formulas. There is an edge between two formulas $\varphi$ and $\psi$ if there is no other formula $\psi'$ satisfying $\varphi \leq \psi' \leq \psi$. This graph is known as the version space~\cite{Mitchell82}.

\figref{conspace} presents part of this graph corresponding to our running example: the bottom  node shows the most specific concept formula, $\varphi_0$, which satisfies all predicates from $S$ satisfied by the initial positive example $(2,2)$. Edges link $\varphi_0$ to more relaxed formulas (formulas with fewer constraints). We use $\varphi_0^{R_1,...,R_n}$ to denote the formula $\varphi_0$ without the predicates $R_1,...,R_n$.

\para{Our Approach}
To learn the concept formula, one has to present examples to the user and prune the inconsistent nodes, until a single node remains. {\alg} performs a guided traversal looking for a path to the concept formula $\varphi_C$. At each step, {\alg} examines a specific node and its immediate neighbours to find a step towards $\varphi_C$.

An immediate neighbour of a node $\varphi$ is $\varphi^R$, the formula $\varphi$ where a single predicate $R$ has been dropped. Note that $\varphi$ logically implies $\varphi^R$.
Examining $\varphi^R$ means checking whether $\varphi_C$ is reachable from it, i.e., whether $R$ is in $\varphi_C$. To check if $R$ is part of $\varphi_C$, \alg\ looks for a distinguishing input between $\varphi$ and $\varphi^{R}$. Since $\varphi$ implies $\varphi^R$, this means finding an example satisfying $\varphi^R \wedge \neg \varphi$.

If an example $e$ such that $e \models \varphi^R \wedge \neg \varphi$ exists, its classification enables progress: if $e$ is a positive example (i.e., $e \models \varphi_C$), \alg\ proceeds towards $\varphi^R$ (and prunes the rest of the space), otherwise, $\varphi_R$'s sub-graph is pruned.

In the special case where predicates are independent, every conjunction of literals over the predicates is satisfiable, and an example $e \models \varphi^R \wedge \neg \varphi$ is guaranteed to exist. Consequently, in this restricted case, every predicate can be classified with a single question. In fact, this is the case handled by classic exact learning algorithms (e.g., \cite[Theorem. 11]{Goldman:1995}).
However, we address the general case, where \textbf{\emph{predicates may be dependent}}, and such distinguishing input might not exist. This leads to the first novel challenge that we address:

\begin{challenge}
When $\varphi^R \wedge \neg \varphi$ is not satisfiable, how can one obtain alternative distinguishing inputs that enable to classify $R$?
\end{challenge}

A na\"ive solution to this challenge is to examine every child of $\varphi^R$: if all have distinguishing inputs with $\varphi$ and one of these inputs is a positive example (satisfies $\varphi_C$), the traversal proceeds towards this child and prunes the rest of the space; otherwise, if all inputs are negative examples, $\varphi^R$'s sub-graph is pruned. However, it is not guaranteed that the children necessarily have distinguishing inputs, in which case \emph{their} children must be examined similarly. While this solution is correct, it is wasteful in the number of questions. This leads us to the second challenge we address:

\begin{challenge}
How can one obtain \textbf{\emph{a minimal number}} of alternative distinguishing inputs?
\end{challenge}

To present a minimal number of questions, we show that instead of examining all children of $\varphi_R$, it suffices to examine a \emph{subset} of children. This subset is the set of predicates in $\varphi_R$ ``preventing'' distinguishing inputs with $\varphi$. That is, predicates preventing the formula $\varphi_R \wedge \neg \varphi$ from being satisfiable. Such predicates are known as \emph{the unsat core} of the formula.

We prove that if there is no distinguishing input for $\varphi_R$ and $\varphi$, it suffices to compute an unsat core of the above formula and consider only the children belonging to the unsat core. If the computed unsat cores are guaranteed to be minimal, we prove that a minimal number of questions is presented. Though finding minimal unsat cores in general theories is EXPSPACE-complete, there are approaches to compute small unsat cores (e.g.,~\cite{CimattiGS11}), and in some theories (such as the ones exemplified in this work), minimal unsat cores can be computed.

\subsection{A Running Example}
We now demonstrate {\alg} on the concept defined in~\secref{overview1}.
Given the initial user-provided example  $(2,2)$, \alg\ first computes the most strict consistent concept, $\varphi_0$ (which implies every consistent concept), which is $\bigwedge_{l \in S_0=\{(2,2) \models R \mid R\in S\}} l$, that is:
\[
\begin{array}{lll}
\varphi_0 & = & \rangep{x}{0}{2} \land \rangep{x}{1}{3} \land\rangep{x}{2}{4}\land \\
&&\rangep{y}{0}{2} \land\rangep{y}{1}{3} \land\rangep{y}{2}{4} \land (x = y)
\end{array}
\]


\para{Distinguishing Inputs} After constructing $\varphi_0$, \alg\ looks for a predicate $R$ that can be classified with a single example. Unfortunately, none of its immediate neighbours in the concept graph has a distinguishing input with $\varphi_0$. For example, for
  $R=\rangep{x}{0}{2}$ there is no distinguishing input, because such an input has to satisfy the (unsatisfiable) formula, $\varphi_0^R \wedge \neg \varphi_0$, which is simplified to:

\[
\begin{array}{ll}
\psi_{S_0}^{\rangep{x}{0}{2}}&\hspace{-0.3cm}\triangleq \bigwedge_{l \in S_0 \setminus \{\rangep{x}{0}{2}\}} l \land \neg \rangep{x}{0}{2} =\\
&\hspace{-0.3cm}\neg \rangep{x}{0}{2} \land \rangep{x}{1}{3} \land \rangep{x}{2}{4}\land \\
&\hspace{-0.3cm}\rangep{y}{0}{2} \land \rangep{y}{1}{3} \land\rangep{y}{2}{4} \land (x = y)
\end{array}
\]

In the following, we use the notation $\psi_{Qs}^{Rs}$ to refer to the formula satisfying the predicates in $Qs$ and not in $Rs$ and the negations of the predicates in $Rs$, that is:
  $\psi_{Qs}^{Rs}=\bigwedge_{l \in Qs \setminus Rs} l \land \bigwedge_{l \in Rs} \neg l$.

Back to our example, the formula $\psi_{S_0}^{\rangep{x}{0}{2}}$ is unsatisfiable due to the dependency between the predicates $\rangep{x}{0}{2}$, $\rangep{y}{0}{2}$, and $x=y$. If there were such examples they would satisfy that $x$ is greater than $2$, $y$ is at most $2$, and $x$ equals $y$, which clearly cannot be satisfied together.

\para{Finding Alternative Distinguishing Inputs}
To find alternative satisfiable formulas, {\alg} computes an \emph{unsat core} of the above formula. For each predicate $R'$ in the unsat core (except for the one at hand, $R$), \alg\ constructs a formula that negates $R'$, in addition to $R$.
If some of these formulas are still unsatisfiable, \alg\ repeats this process, computes a (new) unsat core, and generates a set of formulas from it.
Finally, \alg\ presents the user an example for each of these formulas.
In our example, \alg\ computes an unsat core of ${\psi}_{S_0}^{\rangep{x}{0}{2}}$, which is $\{\rangep{x}{0}{2},\rangep{y}{0}{2},x=y\}$ and generates the formulas
${\psi}_{S_0}^{\rangep{x}{0}{2}, (x=y)}, {\psi}_{S_0}^{\rangep{x}{0}{2}, \rangep{y}{0}{2}}$. These formulas are satisfiable by $(3,2)$ and $(3,3)$ respectively. Therefore, {\alg} presents these points to the user.

\para{Inferring Classifications from the Alternative Formulas}
If one of these examples, corresponding to $\psi^{R,R_1,...,R_k}_{S_0}$, is classified by the user as a positive example, then none of the negated predicates is part of $\varphi_C$, and thus $R,R_1,...,R_k$ are dropped from the current formula. However, if all of them are negative, then it is only guaranteed that the predicate at hand, $R$, is part of the correct concept $\varphi_C$.
For example, in our example, both points $(3,3)$ and $(3, 2)$ are negative, and thus \alg\ infers that $\rangep{x}{0}{2}$ is part of $\varphi_C$. Note that although these formulas negate additional predicates, $x=y$ and $\rangep{y}{0}{2}$, these cannot be classified at this point. Indeed, eventually $x=y$ will be dropped, while $\rangep{y}{0}{2}$ will be part of $\varphi_C$.
However, the next step of \alg, which considers the predicate $\rangep{x}{2}{4}$, infers differently. As before, there is no distinguishing input for $\varphi_0$ and $\varphi_0^{\rangep{x}{2}{4}}$ (i.e., ${\psi}^{\rangep{x}{2}{4}}_{S_0}$ is unsatisfiable). Therefore, {\alg} considers the unsat core $\{\rangep{x}{2}{4},\rangep{y}{2}{4},x=y\}$ and generates the relaxed formulas ${\psi}_{S_0}^{\rangep{x}{2}{4},x=y}$ and ${\psi}_{S_0}^{\rangep{x}{2}{4},\rangep{y}{2}{4}}$. Both formulas are satisfiable, by $(1,2)$ and $(1,1)$, and both are positive. However, this time after the first example, $(1,2)$, is presented to the user, {\alg} infers immediately (without presenting $(1,1)$) that $\rangep{x}{2}{4}$ \emph{and} $x=y$ are not in $\varphi_C$, and thus it updates the current candidate formula to $\varphi_0^{\rangep{x}{2}{4},x=y}$.
In the next step, \alg\ looks for distinguishing inputs from the new candidate formula, and so it constructs $\psi_{S_0 \setminus \{{\rangep{x}{2}{4},x=y}\}}^{\rangep{y}{2}{4}}$, namely it ignores the predicates $\rangep{x}{2}{4},x=y$, as they no longer affect the classifications.
We note that in fact $\rangep{x}{2}{4}$ is implied by the predicate $\rangep{x}{0}{2}$, and thus is classified as redundant by \alg\ immediately after learning $\rangep{x}{0}{2}$ -- we ignore this step here to exemplify how \alg\ classifies a predicate as not part of $\varphi_C$.

\subsection{\alg\ Extensions}
We use the logic described for learning formulas over conjunctions, to learn other concept classes: disjunctions and DNFs.

\para{D-\alg} The disjunctive variation of \alg\ is dual to the conjunctive.
While the conjunctive variation, C-\alg, generalizes from the positive examples and learns which constraints must be met by examples in the concept, D-\alg\ generalizes from the \emph{negative} examples, and learns which constraints eliminate examples from being part of the concept.
\figref{conspace2} visually demonstrates the difference between the classes:
C-\alg\ learns a consecutive region in the concept space that contains all positive examples, while D-\alg\ learns the same region only for the negative examples.

\sloppy
\para{Gen-\alg} Gen-\alg\ learns more complex formulas that can capture general concepts, in which there is no single consecutive region for the positive examples or the negative examples (as illustrated in~\figref{conspace2}). Ideally, to learn such concepts, Gen-\alg\ would simply invoke C-\alg\ to learn a conjunction for each region (independently) and then return the disjunction over these conjunctions.
However, there are two main issues with this approach that Gen-\alg\ has to address:
\begin{itemize}[nosep,nolistsep]
  \item How to guarantee that every region has been covered? It cannot assume that the user provides enough examples.
  \item How to handle intersecting regions? C-\alg\ may over-generalize such regions, resulting in an incorrect specification.
\end{itemize}

To address the first challenge, Gen-\alg\ maintains two formulas:
 \begin{inparaenum}[(i)]
   \item $\varphi_P$, satisfied by the positive examples, and
 \item $\varphi_N$, satisfied by the negative examples.
 \end{inparaenum} While there is an example not satisfying any of them, Gen-\alg\ asks the user for the example's classification, and accordingly adds a conjunction to $\varphi_P$ or $\varphi_N$.

To address the second challenge, we first identify the pitfall of employing C-\alg\ as-is:
C-\alg\ relies on the fact that every example ``outside'' of the (single) region is classified as negative example. However, this is not true for Gen-\alg\ as examples ``outside'' of a certain region may be classified as positive if they belong to a different region. Since C-\alg\ generates examples that are ``close'' the the current candidate hypothesis, it can learn regions that are ``sufficiently apart'' from others. For regions that are ``too close'' or even intersect, over-generalization may still occur in two cases:
\begin{inparaenum}[(i)]
  \item if an example is in the intersection of two regions, and
  \item if several examples are generated to classify a predicate $R$ (due to dependency between predicates), which leads to removing $R$ if one of them is positive.
\end{inparaenum}
In the first case, C-\alg\ will not be able to isolate the regions and will return an over-generalization containing them, and thus Gen-\alg\ has to detect this and ignore the conjunction. In the second case, Gen-\alg\ avoids over-generalization by modifying C-\alg\ to examine \emph{all} examples and 
eliminate the negative examples with a \emph{disjunction} for each negative example. While this results in a formula which is not a DNF (as its conjunction may be over disjunctions and not only literals), it can be easily transformed to a DNF, and thus we refer to learning such formulas as learning a DNF. We provide further details in~\secref{dnfspex}.

\begin{figure}
\vspace{0.1cm}
\centering
  \includegraphics[page=1,height=4.2cm,width=7cm,clip=true,trim=0pt 0pt 600pt 0pt]{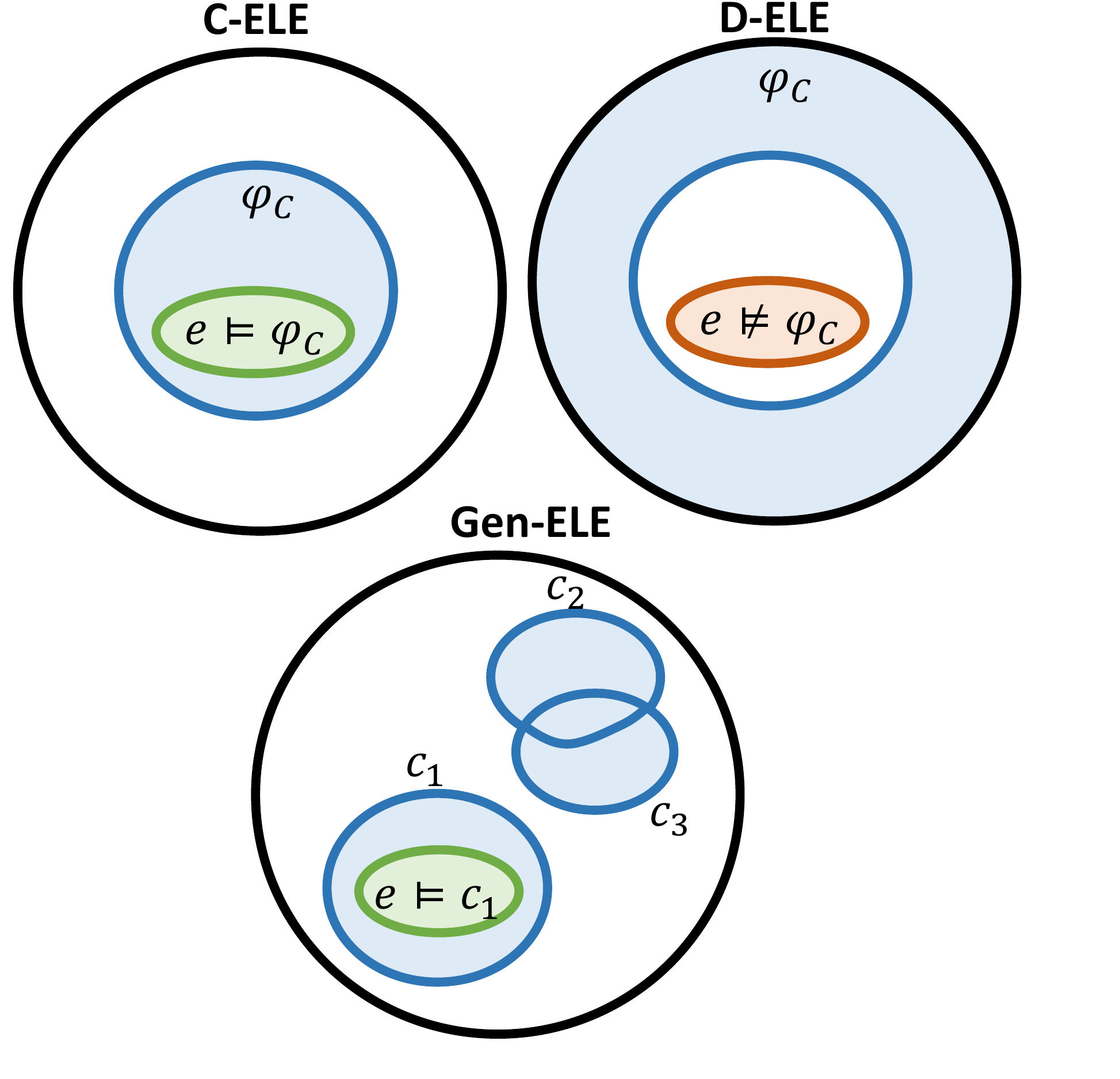}
  \caption{Illustration of the concept spaces.}
  \figlabel{conspace2}
\end{figure} 
\section{Exact Learning from Examples}\seclabel{spex}

In this section, we define formally the problem of learning an exact specification from examples. 

\para{Specifications} We consider three types of specifications:
\begin{inparaenum}[(i)]
  \item DNF specifications where the formula is in disjunctive normal form,
  \item conjunctive specifications, a restricted case of DNF where there is a single cube (i.e., there are no disjunctions), and
  \item disjunctive specifications, a restricted case of DNF where each cube contains a single literal (i.e., there are no conjunctions).
\end{inparaenum}
The specifications are defined over arbitrary predicates, defined over the example domain.
We next formally define them.

\begin{definition}[A DNF Specification]\deflabel{conjunctivespec}
Let the example domain $D$ be a set and $S$ be a set of predicates over $D$, namely $\forall R\in S$, $\exists n\in \mathbb{N}: R\subseteq D^n$.
A \emph{DNF specification} is a formula $\varphi(d)~=~\bigvee [\bigwedge_{l\in A_i} l(d)]$, where each $A_i$ is a subset of literals over $S$, that is $A_i \subseteq \{R,\neg R \mid R \in S\}$.
\end{definition}

\begin{definition}[A Conjunctive Specification]\deflabel{conjunctivespec}
A \emph{conjunctive specification} is a DNF specification with a single cube, namely
$\varphi(d)=\bigwedge_{l\in A} l(d)$ where $A\subseteq \{R,\neg R \mid R \in S\}$.
\end{definition}

\begin{definition}[A Disjunctive Specification]\deflabel{disjunctivespec}
A \emph{disjunctive specification} is a DNF specification where each cube has a single literal, namely
$\varphi(d)=\bigvee_{l\in A} l(d)$ where $A\subseteq \{R,\neg R \mid R \in S\}$.
\end{definition}

\para{Exact Learning of Specifications}
We address the problem of exact learning of specifications. In exact learning, the goal is to precisely classify \emph{every} example in the domain $D$, without necessarily explicitly seeing every input-output example. We consider the teacher-student model where the student (i.e., the algorithm) can ask the teacher (i.e., the oracle or user) only membership questions, that is ask for the output of a given input. We further allow the teacher to provide some initial positive and/or negative examples, however the teacher need not provide examples, and in any case the student obtains the examples it needs by interacting with the teacher.
We next formally state this (interactive) learning problem.

\begin{definition}[Exact Learning from Examples (\name)]\sloppy
Let $D$ be a domain, $S$ be a set of predicates over $D$, $\varphi_C$ be an unknown specification over $S$ (to be discovered), $E_P,E_N\subseteq D$ be initial sets of positive and negative examples (i.e., $\forall d\in E_P.\varphi_C(d)$ and $\forall d\in E_N.\neg\varphi_C(d)$), and an oracle that can precisely classify any example in $D$. The goal of exact learning from examples is  to learn a specification $\varphi$ over $S$ such that: $\forall d \in D. \varphi_C(d) \leftrightarrow \varphi(d)$.
\end{definition}

 We refer to the above problem as a C-\name, D-\name, or DNF-\name\ if the specification is
 conjunctive, disjunctive, or DNF (resp.).

\para{\name's Complexity Class} \name\ was extensively studied and in particular it was shown to be EXPTIME for the special case where the domain is a set of boolean vectors and the predicates are monomials over the vectors~\cite{Angluin:1993}.
This implies that our general setting of DNF-\name, which does not restrict the domain or the predicates, is also EXPTIME.
The work of~\cite{Angluin:1993} also implies that D-\name\ is EXPTIME since it can be seen as a special setting of D-\name\ where the domain is the boolean vectors and $S$ contains conjunctions.
To show that C-\name\ is also EXPTIME we prove the following claim: 

\begin{claim}\label{ICSIPE-lowerbound}
Let
\begin{itemize}[nosep,nolistsep]
  \item $D$ be a set of boolean vectors: $\{(x_0,...,x_k)\mid \forall i. x_i \in \{0, 1\}\}$.
  \item $S$ be a set of disjunctions: $\{(x_0 \vee x_j), (x_0 \vee \neg x_j) \mid 1\leq j \leq k\}$.
  \item $E_P = E_N = \emptyset$.
\end{itemize}
For any \name\ algorithm there is a conjunctive specification $\varphi_C$ which presents $\Omega(2^{|S|})$ membership queries to the oracle.
\end{claim}
Intuitively, the specification is a CNF, which here is equivalent to learning DNF.
Proof is provided in~\appref{proofs}.

\section{The C-SPEX and D-SPEX Algorithms}\seclabel{class}
In this section, we present our exact learning algorithm for the restricted classes of conjunctive and disjunctive specifications. We begin with the high-level algorithm, then show the algorithm itself and instantiate it to the C-\name\ and D-\name\ algorithms, and finally prove that {C-\alg} and D-\alg\ generate a minimal number of examples and that for a useful class this number is linear.

\subsection{C-\alg\ and D-\alg\ in a Nutshell}\seclabel{spex:nutshell}
In this section, we present the pseudo code of \alg\ and discuss the main differences between its conjunctive and disjunctive variations.

\para{The Guided Traversal}
\algref{pseu2} shows the pseudo code of \alg's guided traversal. The algorithm takes as arguments the set of predicates $S$ and the initial set of positive and negative examples $E_P$ and $E_N$ (which may be empty). It begins by constructing the most specific formula and storing its literals in $S_0$, which provides the ``alphabet'' of the concept formula $\varphi_C$ to be learned.  

The guided traversal classifies each literal in $S_0$ as part of $\varphi_C$ or not. It maintains two sets, $S_P$ and $S_N$, storing the predicates classified so far as part of $\varphi_C$ ($S_P$) or not ($S_N$).
\alg\ iteratively classifies literals until all are at $S_P$ or $S_N$. At each step, it invokes a function that returns a literal, which minimizes the number of examples needed for classification, and its classifying examples. It then gradually asks the oracle for their output, until it can classify the literal. If the literal was classified to $S_P$, literals that are implied by $S_P$ and the new literal are classified to $S_P$, too.
Finally, \alg\ generates $\varphi_C$ by constructing a conjunctive or disjunctive formula from $S_P$, cleans it by removing redundant literals, and returns it.

\begin{algorithm}[t]
\small
\DontPrintSemicolon
\LinesNumbered
$S_0$ = get literal set of the most specific hypothesis from $S$, $E_P$, $E_N$\;
$S_P$ = $S_N$ = $\emptyset$\;
\While{$S_P \cup S_N \subsetneq S_0$}{
    $l$, exs = find a literal that requires a minimal number of examples\;
    Get feedback on $exs$ until $l$ can be classified to $S_P$ or $S_N$\;
    \lIf{$l\in S_P$}{
        add to $S_P$ all literals $\hat{l}$ implied by $S_P$ and $l$
    }
}
Construct $\varphi_C$ by collecting all the literals in $S_P$\;
Clean $\varphi_C$ by removing implied literals\;
\Return{$\varphi_C$}
\caption{SPEX Pseudo Code($S$, $E_P$, $E_N$)}\label{Alg:pseu2}
\end{algorithm} 
\para{C-\alg\ and D-\alg} The above pseudo code is the framework of variations the C-\alg\ and D-\alg. While the framework is identical, the two variations are not identical but dual: C-\alg\ generalizes from positive examples, whereas D-\alg\ generalizes from negative examples. We next informally present the differences:
\begin{itemize}[nosep,nolistsep]
  \item Initialization: both variations initialize $S_0$ such that the conjunction (in C-\alg) or disjunction (in D-\alg) over its elements implies $\varphi_C$.
  In C-\alg, $S_0$ contains the literals from $S$ satisfied by all positive examples.
   In D-\alg, $S_0$ contains the negations of literals from $S$ satisfied by all negative examples.
  \item Constructing examples:
  in both variations, the goal is to learn which literals from $S_0$ are part of $\varphi_C$, and thus to classify the literals, \alg\ constructs distinguishing inputs, however those are constructed differently.
  In C-\alg, to infer whether a literal $l$ in $S_0$ is in $\varphi_C$, a distinguishing example for the conjunction over $S_0 \setminus S_N$ (the most strict hypothesis consistent with the current positive examples) and the same hypothesis only without $l$ satisfies all the literals in $S_0 \setminus S_N$ (but $l$) and $\neg l$. If such example is positive, $l$ is not in $\varphi_C$, otherwise it is. Intuitively, correctness follows because if the example $e$ is positive, i.e., $e \models \varphi_C$, but does not satisfy $l$, i.e., $e \not \models l$, $l$ cannot be part of the conjunction $\varphi_C$. In contrast, in D-\alg, to infer whether a literal $\neg l$ is in $\varphi_C$, a distinguishing example satisfies $\neg l$ and none of the other literals in $S \setminus S_N$. If such example is negative, $\neg l$ is not in $\varphi_C$, otherwise it is. Intuitively, correctness follows because if the example $e$ is negative, i.e., $e \not \models \varphi_C$, then if $\varphi_C$ would have contained $\neg l$, $e$ should have been a positive example since it satisfies the disjunction.
      We note that in case there are no such distinguishing inputs, \alg\ considers alternative formulas whose distinguishing inputs enable to infer the classification similarly (as will be described later).
  \item Implications: in C-\alg, if a literal $l$ is added to $S_P$, any other literal implied by $S_P \cup \{l\}$ is also in $S_P$ (since any positive example satisfies $S_P$ and $l$, and thus this literal). In D-\alg, if a literal $\neg l$ is added to $S_P$, any other literal that is implied by the disjunction is added to $S_P$ (since there are no positive examples satisfying the disjunction but not this literal) and any other literal that implies the disjunction is added to $S_P$ (since there are no positive examples satisfying this literal but not the disjunction and adding it to the disjunction does not strengthen or relaxes the disjunction).
      Removing the implied literals is only required to complete the classification of each literal in $S_0$, and these literals do not affect the final formula $\varphi_C$, since it is cleaned from redundant literals.
\end{itemize}

\begin{algorithm}[t]
\small
\DontPrintSemicolon
\LinesNumbered
 $S_0$ = init($S$, [isCon? $E_P$ : $E_N$]) \;\lnlabel{prediter:sinit}
 $S_{P}$ = $\emptyset$ ;  $S_{N}$ = $\emptyset$ \;
 \While {\textsc{(}$S_{P} \cup S_{N} \subsetneq S_{0}$\textsc{)}}{ \lnlabel{learnLoopBegin}
    ($l$, $Exs$) = \text{getMinLiteralNExamples}($S_{0}$,$S_{P}$,$S_{N}$,form)\; \lnlabel{nextPred}
   classified = false\;
    \For {$(Rs,\hat{e}) \in Exs$}{ \lnlabel{prediter:examBegin}
        \lIf{askUser($\hat{e}$, $E_P$, $E_N$)$=$(isCon? neg : pos)}{ continue}\lnlabel{prediter:ask} 
        $S_{N} = S_{N} \cup Rs$\lnlabel{learnPos}\;
        classified = true\; break\;%
    }\lnlabel{prediter:examEnd}
    \lIf{!classified}{
        $S_{P} = S_{P} \cup implied($$S_P$,$l$)   \lnlabel{nextpredImplied}
    }
    }
 \lnlabel{learnLoopEnd}
 $\varphi_C$ = isCon? $\bigwedge_{l \in S_P} l$ : $\bigvee_{l \in S_P} l$\;\lnlabel{removeImplied1Begin}
  \For {$l \in \varphi_{C}$}{
            \lIf{$\varphi_{C} \setminus \{l\} \models l$}{
                $\varphi_{C} = \varphi_{C} \setminus \{l\}$
           }
       }      \lnlabel{removeImplied1End}

\Return{$\varphi$}

\caption{SPEX($S$, $E_P$, $E_N$, isCon)}\label{Alg:prediter}
\end{algorithm}

\subsection{The SPEX Algorithm}
In this section, we present the actual algorithm of \alg, shown in \algref{prediter}, which instantiates the template of \algref{pseu2}. 
{\alg} takes as arguments the set of predicates $S$, the initial sets $E_P$ and $E_N$ of positive and negative examples, and the \emph{isCon} flag indicating whether to learn a conjunctive or a disjunctive specification. To instantiate it, two operations are required, \scode{init} and \scode{implied}, implemented differently by C-\alg\ and D-\alg.

\alg\ begins with initializing $S_0$ to the set of all possible literals that $\varphi_C$ may contain (using \scode{init}) and 
the two literal sets, $S_P$ and $S_N$, to the empty sets. 
 Then, \alg\ iteratively generates examples to classify literals in $S_{0}$ until all are in $S_P$ or $S_N$ (Lines~\slnref{learnLoopBegin}--\slnref{learnLoopEnd}). At each iteration, \alg\ invokes \scode{getMinLiteralNExamples} that picks the next literal to classify $l$ and returns $l$ along with the examples that imply its classification. Each example is accompanied with a set of literals $Rs$ containing the \emph{relaxed} literals (in C-\alg\ this means literals that are negated, and in D-\alg\ this means literals that are not negated).
Then, \alg\ gradually iterates the examples to classify $l$ (Lines~\slnref{prediter:examBegin}--\slnref{prediter:examEnd}). First, it obtains their output (Line~\slnref{prediter:ask}) using \scode{askUser} (whose code is omitted) that gets the example's classification (positive or negative) either from the available examples or, if the example is new, from the oracle and adds the new example to $E_P$ or $E_N$ accordingly.
After obtaining the output, \alg\ classifies according to its duality. 
In C-\alg, if the example is positive, this indicates that $l$ and the rest of the literals in $Rs$ (which includes $l$) are not in $\varphi_C$, and thus all are added to $S_N$, and \alg\ continues to classify the next literal.
Otherwise, if all the examples are negative, this indicates that $l$ is in $\varphi_C$, and thus the set of literals implied by $S_P$ and $l$ (which includes $l$) is computed using \scode{implied} and added to $S_P$ (Line~\slnref{nextpredImplied}).
D-\alg\ is dual: it adds $l$ and $Rs$ to $S_N$ if one of the examples is negative, or adds $l$ to $S_P$ if all the examples are positive.
Finally, \alg\ generates $\varphi_C$ from $S_P$ and cleans it by removing implied literals (Lines~\slnref{removeImplied1Begin}--\slnref{removeImplied1End}).

\subsection{Computing the Next Literal to Classify}
In this section, we present \scode{getMinLiteralNExamples} (\algref{exam}), abbreviated to \scode{getMin}, that picks the next literal to classify and returns it along with the examples implying its classification. Each example is accompanied with the set of its relaxed literals. 
 We first describe how to compute for a given literal a minimal set of examples that imply its classification, and then describe \scode{getMin} that finds a literal whose example set is of minimal size.

\begin{table*}\centering
\ra{1.2}
\begin{tabular}{p{2cm}|p{3.8cm}|p{6.3cm}|p{3.8cm}}\toprule
variation & init($S$, $E_P$, $E_N$) & implied($S_P$, $l$)& $\psi^{Rs}_{Os}$  \\
\midrule
C-\alg & $\{l \in S \mid \forall e\in E_P. e\models l \}$ & $\{\hat{l} \mid \bigwedge_{q \in S_P \cup \{l\}} q  \models \hat{l}\}$& $\bigwedge_{Q\in Rs} \neg Q \wedge \bigwedge_{Q\in Os \setminus Rs} Q$ \\
D-\alg & $\{\neg l \in S \mid \forall e \in E_N. e \models l\}$ & $\{\hat{l}\mid \bigvee_{q \in S_P \cup \{l\}} q \models \hat{l}\}\cup \{\hat{l} \mid \hat{l} \models \bigvee_{q \in S_P \cup \{l\}} q \}$ &$\bigwedge_{Q\in Rs} Q \wedge \bigwedge_{Q\in Os \setminus Rs} \neg Q$ \\
\bottomrule
\end{tabular}
\caption{The template functions of the two \alg\ variations.}\tablabel{funcs}
\end{table*}

 \para{Computing the Minimal Example Set of a Literal}
 Ideally, a literal $l$ can be classified using a single example. 
 To check if there is such example, \scode{getMin} constructs a formula $\psi^{Rs}_{S_0 \setminus S_N}$, where $Rs=\{l\}$, ``isolating $l$'s effect'':
 \begin{itemize}[nosep,nolistsep]
   \item In C-\alg, $\psi^{Rs}_{S_0 \setminus S_N}$ is the conjunction of $\neg l$ and the literals in $S_0$, except for $l$ and the literals classified to $S_N$.
   \item In D-\alg, $\psi^{Rs}_{S_0 \setminus S_N}$ is the conjunction of $l$ and the negations of the other literals in $S_0\setminus S_N$.
 \end{itemize}
 If $\psi^{Rs}_{S_0 \setminus S_N}$ is satisfiable, \emph{any} example in $D$ satisfying it is an example whose classification implies $l$'s classification. If $\psi^{Rs}_{S_0 \setminus S_N}$ is unsatisfiable, but there is a single way to relax $\psi^{Rs}_{S_0 \setminus S_N}$ (by removing specific literals), then similarly any example satisfying the relaxed formula can serve as the single classifying example.

  However, if there are multiple ways to relax $\psi^{Rs}_{S_0 \setminus S_N}$ (e.g., $l_1$ or $l_2$ may be removed from it), \scode{getMin} has to consider every relaxed formula and generate an example for each (it does not necessarily mean that all will be presented to the oracle). To find a minimal number of relaxed formulas, \scode{getMin} uses UNSAT cores (i.e., unsatisfiable sets of literals from the formula) that may be computed from the unsatisfiable formula $\psi^{Rs}_{S_0 \setminus S_N}$, for example using an SMT-solver (e.g.,~\cite{DeMoura:2008}). The UNSAT cores must contain $l$ (because $S_0\setminus S_N$ is satisfiable) and some literals from $S_0 \setminus S_N$ not in $S_P$ (otherwise, $l$ is implied from $S_P$, but then it would have been removed by \alg\ before invoking \scode{getMin}).
  Each of these literals is a possibility to consider except for $l$ and literals from $S_P$ (as these dominate the examples'
  classification, regardless of $l$'s classification). Thus, for each \scode{getMin} generates a formula extending $Rs$ with this literal. If the new $\psi^{Rs}_{S_0 \setminus S_N}$ is still unsatisfiable, an additional core is computed and new relaxed formulas replace the former relaxed formula.
      Since the relaxed formulas are uniquely identified by their relaxed literals, \scode{getMin} maintains the relaxed literal set of each relaxed formula, which are stored in $sets$.

\begin{algorithm}[t]
\small
\DontPrintSemicolon
\LinesNumbered
\For{max = 1; ; max++}{\lnlabel{nextpredLoopBegin}
    \For{$l \in S_{0} \setminus (S_{P} \cup S_{N})$}{\lnlabel{nextpredInnerLoopBegin}
        $sets$ = $\{\{l\}\}$\;\lnlabel{qsinit}
        \For{$Rs\in sets$}{ \lnlabel{satLoopBegin}
            \lIf{sat($\psi^{Rs}_{S_0\setminus S_N}$)}{continue}\lnlabel{getexam:check}
            $core$ = $unsatCore(\psi^{Rs}_{S_0\setminus S_N})\setminus (S_P\cup Rs)$\; \lnlabel{getexam:core}\lnlabel{getexam:relaxedBegin}
            $sets$ = $sets \setminus \{Rs\} \cup \{Rs \cup \{Q\} \mid Q \in core\}$\; \lnlabel{getexam:relaxedEnd}
            \lIf{$|sets| > max$}{break}\lnlabel{violation}
        }\lnlabel{satLoopEnd}
        \lIf{$|sets| > max$}{continue}\lnlabel{exceeds}
         \Return{$l,\{(Rs,ex(\psi^{Rs}_{S_0\setminus S_N}))\mid Rs \in sets\}$}
    }
}\lnlabel{nextpredInnerLoopEnd}

\caption{getMinLiteralNExamples($S_{0}$,$S_{P}$,$S_{N}$)}\alglabel{exam}
\end{algorithm}

\para{Computing the Min Literal} To find a literal requiring a minimal number of examples, \scode{getMin} sets a bound on this number with the variable $max$ and increases it only if all literals require more examples (Line \slnref{nextpredLoopBegin}).
After fixing $max$, every unclassified literal $l$ is checked whether it can be classified using at most $max$ examples (Line \slnref{nextpredInnerLoopBegin}). To this end, \scode{getMin} initializes $sets$ to contain the initial $Rs$ set, $\{l\}$, (Lines \slnref{qsinit}) and replaces each literal set whose formula $\psi^{Rs}_{S_0\setminus S_N}$ is unsatisfiable with its relaxed sets, as previously described.
Then, a loop updates $sets$ until:
 \begin{inparaenum}[(i)]
   \item every $Rs$ satisfies that $\psi^{Rs}_{S_0\setminus S_N}$ is satisfiable, or
 \item the size of $sets$ exceeds $max$ (Lines \slnref{satLoopBegin}--\slnref{satLoopEnd}).
 \end{inparaenum}
 To determine whether for a given $Rs$, $\psi^{Rs}_{S_0\setminus S_N}$ is satisfiable, \scode{getMin} uses an SMT-solver (Line \slnref{getexam:check}). If $\psi^{Rs}_{S_0\setminus S_N}$ is unsatisfiable, an UNSAT core is obtained from the SMT-solver (we also reduce it to be minimal by removing redundant literals, we omit this part from the code), and $sets$ is updated to exclude $Rs$ and include all the sets consisting of $Rs$ and a single literal from the unsat core that is not in $S_P$ or $Rs$ (Lines \slnref{getexam:relaxedBegin}--\slnref{getexam:relaxedEnd})\footnote{More precisely, in C-\alg\ the core actually contains \emph{negations} of literals from $Rs$, and in D-\alg\ \emph{negations} of literals from $S_P$, but the core is cleaned from these literals without the negations.}. 
If the extension of $sets$ results in exceeding $max$, the loop terminates (Line~\slnref{violation})
 and the next literal is examined (Line~\slnref{exceeds}). Otherwise, $l$ is returned along with the set of pairs consisting of the $Rs$ sets and their corresponding examples. The examples are obtained from the SMT-solver, denoted \scode{ex($\psi$)}.


\subsection{Implementing C-\alg\ and D-\alg}
In this section, we describe the operations that instantiate C-\alg\ and D-\alg:
\begin{inparaenum}[(i)]
  \item init,
  \item imply, and
  \item $\psi^{Rs}_{Os}$,
\end{inparaenum}
listed in \tabref{funcs}.

\para{C-\alg} This variation implements these operations as follows:
\begin{itemize}[nosep,nolistsep]
  \item \scode{C-init} returns the conjunction over all literals in $S$ satisfied by all positive examples.
   This formula implies the specification $\varphi_C$: literals not in it are not satisfied by one of the positive examples and thus are
   not in~$\varphi_C$.
  \item \scode{C-implied} returns the set of literals implied by the conjunction of $S_P$ and the literal $l$.
  \item \scode{C-$\psi^{Rs}_{Os}$} returns the conjunction of:
  \begin{inparaenum}[(i)]
    \item the negations of the literals in $Rs$, i.e., $l$ and the literals relaxed to obtain a satisfiable formula, and
    \item the other literals in $S\setminus S_N$. Literals in $S_N$ may be determined arbitrarily as they do not affect the classification: \alg\ observed positive and negative examples satisfying them.
  \end{inparaenum}
\end{itemize}

\begin{theorem}\thelabel{conpred}
  Given $D$, $S$, $\varphi_C$ an unknown conjunctive specification over $S$, and initial positive and negative examples $E_P$ and $E_N$. Let C-\alg\ be \alg\ with \scode{C-init}, \scode{C-implied}, and \scode{C-$\psi^{Rs}_{Os}$}.
  C-\alg\ is a C-\name\ algorithm, i.e., it learns a conjunctive specification $\varphi$ over $S$ such that: $\forall d. \varphi_C(d) \leftrightarrow \varphi(d)$.
\end{theorem}
Proof is provided in~\appref{proofs}.

\para{D-\alg} This variation implements the operations as follows:
\begin{itemize}[nosep,nolistsep]
  \item \scode{D-init} returns the disjunction over the negations of literals in $S$ satisfied by all negative examples.
   This formula implies the specification $\varphi_C$: a negation of a literal not in it is satisfied by a negative example and thus is not in~$\varphi_C$.
  \item \scode{D-implied} returns the set of literals implied by or implying the disjunction of $S_P$ and the literal $l$.
  \item \scode{D-$\psi^{Rs}_{Os}$} returns the conjunction of:
  \begin{inparaenum}[(i)]
    \item the literals in $Rs$, checked whether they sufficient to satisfy the disjunction, and
    \item the negations of the literals in $S\setminus S_N$, which include $S_P$ that contains literals known to be sufficient to satisfy the disjunction.
  \end{inparaenum}
\end{itemize}

\begin{theorem}\thelabel{cnfpred}
  Given $D$, $S$, $\varphi_C$ an unknown disjunctive specification over $S$, and initial positive and negative examples $E_P$ and $E_N$. Let D-\alg\ be \alg\ with \scode{D-init}, \scode{D-implied}, and \scode{D-$\psi^{Rs}_{Os}$}.
  D-\alg\ is a D-\name\ algorithm, i.e., it learns a disjunctive specification $\varphi$ over $S$ such that: $\forall d. \varphi_C(d) \leftrightarrow \varphi(d)$.
\end{theorem}
Proof is provided in~\appref{proofs}.

\subsection{Complexity Analysis}
In this section, we present the theorem stating that \alg\ asks the minimal number of questions and characterize when this number is linear. Proofs are in~\appref{proofs}.
\begin{theorem}\thelabel{predcomplex}
  Given $D$, a set of literals $S$ of size $n$, and initial examples $E_P$
   and $E_N$. If C-\alg\ or D-\alg\ present $\Omega(f(n))$ questions for some function $f$, any C-\name\ or D-\name\ algorithms present $\Omega(f(n))$ questions.
\end{theorem}
\sloppy
\begin{theorem}\thelabel{predcomplexlinear}
  If at any iteration of C-\alg\ or D-\alg\ there is a literal $l\in S_{0}$ such that  
  \scode{C-$\psi^{Rs}_{Os}$} or \scode{D-$\psi^{Rs}_{Os}$} (resp.) are satisfiable,
   C-\alg\ and D-\alg\ complete in a linear number of questions.
\end{theorem}

Intuitively, if this condition is satisfied, at each invocation of \scode{getMinLiteralNExamples} there is a literal $l$ for which the formula $\psi^{\{l\}}_{S\setminus S_N}$ is satisfiable, and thus a single example is generated. This bounds the number of examples to $|S_0|$, namely linear.

\subsubsection{Classes Learned with a Linear Number of Questions}
This section focuses on a useful class of predicates: predicates that pertain only to the binary relative comparison of values $x,y$, i.e., $x<y,x\leq y, x = y, x\neq y$. For this class, the conditions of \theref{predcomplexlinear} are satisfied for C-\alg, namely it learns with a linear number of questions.

 \begin{claim}\label{relativecomp}
   If $S$ consists of binary relative comparison predicates only, any concept is learned with a linear number of questions.
 \end{claim}
Intuitively, this holds since at each step C-\alg\ picks the points $x,y$ that are closest. When C-\alg\ negates their relation, the only relations that are affected are the ones pertaining to points equal to $x$ or $y$. However, there is only a single possibility to relax these relations (all have to be negated), which results in a single example to consider.
Proof is provided in \appref{proofs}.

Note that the user need not be aware of this condition nor prove it-- \alg\ is self-adaptable and in particular if it is possible to learn with a linear number of questions, \alg\ will discover this during the execution. 
\section{Gen-SPEX}\seclabel{dnfspex}
In this section, we present the Gen-\alg\ algorithm, which enables to learn arbitrarily complex specifications, where positive and negative examples do not necessarily satisfy the same constraints. To learn such specifications, Gen-\alg\ learns sets of constraints and joins them at the end with a disjunction, which forms the desired specification. To enable capturing any specification, Gen-\alg\ does not assume that the initially provided examples (in $E_P$ and $E_N$) satisfy the same constraints and thus generalizes each separately. However, it may happen that during the learning, some examples are discovered as satisfying the same set of constraints.

To guarantee that no set of constraints is missed, Gen-\alg\ learns two sets of constraints, for the positive examples and for the negative examples. When the sets (combined) cover every example, it is guaranteed that no set of constraints could have been missed. This is implemented by Gen-\alg, described in \secref{genalg}.

To learn a set of constraints, which is a ``sub-concept'' added to the desired concept, Gen-\alg\ invokes a slightly modified version of C-\alg. The modification is required because C-\alg\ learns a conjunction and thus assumes that every example in the concept is positive and any other example is negative.
However, Gen-\alg\ uses C-\alg\ to learn ``sub-concepts'', and thus, C-\alg\ may no longer rely on this assumption: positive examples may now be part of a different ``sub-concept'', even though from C-\alg's perspective they should have been classified as negative. As a result, only negative examples provide a guaranteed classification of literals, and this leads to two modifications. First, a literal is classified to $S_N$ only if \emph{all} examples returned by \scode{getMinLiteralNExamples} are positive (and not just a single one). Even then, literals in $S_N$ are not literals which are guaranteed to be excluded from the final formula, but rather literals satisfying that if the literals in $S \setminus S_N$ are satisfied, then they need not be satisfied (and thus at the end, when $S_P=S\setminus S_N$, they can be ignored). This leads to the second modification: the formulas $\psi$ (in \scode{getMinLiteralNExamples}) do not ignore literals in $S_N$.
Further details are in \secref{modifiedc}.

The second modification results in over-generalizing examples belonging to two (or more) sub-concepts. This is because such examples satisfy the literals of both concepts, and thus when C-\alg\ negates the literals of the first concept, the other concept's literals are satisfied, resulting in observing only positive examples and thus adding these literals to $S_N$. When C-\alg\ negates the literals of the second concept, the literals of the first concept are satisfied (because the second modification ensures that the literals in $S_N$ are satisfied), and thus the second concept's literals are also added to $S_N$. To exclude over-generalizing conjunctions, Gen-\alg\ invokes the \scode{overgen} operation, described in \secref{overgen}.

\begin{algorithm}[t]
\small
\DontPrintSemicolon
\LinesNumbered
$C_P$ = $\emptyset$;
$C_N$ = $\emptyset$\;
\For{$e \in E_P$}{ \lnlabel{genalg:initBegin}
    \lIf{$e \models \bigvee_{con \in C_P} con$}{continue}
     con = C-SPEX(e, $E_P$, $E_N$, true)\;
     \lIf{!overgen(con, e, $E_P$, $E_N$, true)}{$C_P$ = $C_P \cup \{con\}$}
}
\For{$e \in E_N$}{
    \lIf{$e \models \bigvee_{con \in C_N} con$}{continue}
     con = C-SPEX(e, $E_N$, $E_P$, true)\;
     \lIf{!overgen(con, e, $E_P$, $E_N$, false)}{$C_N$ = $C_N \cup \{con\}$}
}\lnlabel{genalg:initEnd}

\While{sat($\neg \bigvee_{con \in C_P} con \land \neg \bigvee_{con \in C_N} con$)}{\lnlabel{genalg:extendBegin}
    e = ex($\neg \bigvee_{con \in C_P} con \land \neg \bigvee_{con \in C_N} con$)\;
    \If{askUser(e, $E_P$, $E_N$) == pos}{
    con = C-SPEX(e, $E_P$, $E_N$, true)\;
     \lIf{!overgen(con, e, $E_P$, $E_N$, true)}{$C_P$ = $C_P \cup \{con\}$}
    } \Else {
    con = C-SPEX(e, $E_N$, $E_P$, true)\;
     \lIf{!overgen(con, e, $E_P$, $E_N$, false)}{$C_N$=$C_N \cup \{con\}$}
    }
}\lnlabel{genalg:extendEnd}
 \For {$con' \in C_P$}{\lnlabel{removeImplied1Begin2}
    \lIf{$\{con'\} \models \bigvee_{con \in C_P \setminus {con'}} \{con\}$}{
        $C_P$=$C_P \setminus \{con'\}$
           }
       }      \lnlabel{removeImplied1End2}
\Return{$\bigvee_{con \in C_P} con$}
\caption{Gen-SPEX($S$, $E_P$, $E_N$)}\label{Alg:cnfspex}
\end{algorithm}

\subsection{The Gen-\alg\ Algorithm}\seclabel{genalg}
Gen-\alg\ (\algref{cnfspex}) learns two formulas, one that generalizes the positive examples and the other that generalizes the negative examples. Each of these formulas is a disjunction over a set of conjunctions capturing a single region. The conjunctions are over literals or disjunctions of literals, and they are learned using a slightly modified version of C-\alg\ (described in~\secref{modifiedc}). 

Gen-\alg\ maintains the formulas' set of conjunctions, stored in $C_P$ and $C_N$, which are initially empty. It begins by examining the initially provided examples in $E_P$ and $E_N$ and while they contain examples not satisfied by any of the conjunctions in $C_P$ or $C_N$, it invokes C-\alg, checks if the resulted conjunction is an over-generalization (using \scode{overgen}), and if not, adds the conjunction to $C_P$ or $C_N$, respectively (Lines~\slnref{genalg:initBegin}--\slnref{genalg:initEnd}).
 We note that since $C_N$ is satisfied by the \emph{negative} examples, C-\alg\ switches the user's classifications when generalizing a negative example, and it is invoked with $E_N$ as $E_P$ and $E_P$ as $E_N$.

 Then, while there is an example not satisfied by any of the conjunctions, Gen-\alg\ obtains such example from the SMT-solver, asks the user for its classification, invokes C-\alg, and, if the resulted conjunction is not an over-generalization, adds the new conjunction to $C_P$ or $C_N$ (Lines~\slnref{genalg:extendBegin}--\slnref{genalg:extendEnd}). 

After the loop terminates, the specification is the disjunction over the conjunctions in $C_P$. Before returning it, the conjunctions are cleaned from redundant ones, which are the ones implying the disjunction of the other conjunctions (Lines~\slnref{removeImplied1Begin2}--\slnref{removeImplied1End2}).

\subsection{The Modifications to \alg}\seclabel{modifiedc}
In this section, we describe the two modifications to \alg\ and \scode{getMinLiteralNExamples}.

\para{Modifications to \alg} As discussed at the beginning of this section, \alg\ is modified to classify literals to $S_N$ only if all examples returned by \scode{getMinLiteralNExamples} are positive. While literals may be classified to $S_P$ when a negative example is observed, this is an over-strict classification, since it excludes positive examples that do not satisfy this literal. Though some of these positive examples may be part of a different sub-concept (and thus will be covered later), others may be part of this concept, and excluding them will cause to splitting this sub-concept into two sub-concepts, which will introduce more questions. To avoid this, Gen-\alg\ excludes only the negative examples by adding to $S_P$ a disjunction for every negative example, defined over the literals in $Rs$. The disjunctions exclude the examples since the examples satisfy the \emph{negations} of literals in $Rs$. Additional literals from $S$ cannot be added to the disjunction because they may be part of the final conjunction.
The code snippet below shows these modifications.

{
\setlength{\interspacetitleruled}{0pt}%
\setlength{\algotitleheightrule}{0pt}%
\removelatexerror
\begin{algorithm}[H]
\setcounter{AlgoLine}{5}
\small
\DontPrintSemicolon
\LinesNumbered
\For {$(Rs,\hat{e}) \in Exs$}{ \lnlabel{prediter:examBegin}
    \lIf{askUser($\hat{e}$, $E_P$, $E_N$)=isCon? pos: neg}{ continue}
    $dis = \bigvee_{l' \in Rs} l'$\;
    $S_{P}$ = $S_{P} \cup \{dis\} \cup implied(S_P, dis)$\;
    classified = true\;
}
    \lIf{!classified}{ $S_{N} = S_{N} \cup \{l\} $}
 \end{algorithm}
}

\para{Modifications to \scode{getMinLiteralNExamples}} As discussed at the beginning of this section, \scode{getMinLiteralNExamples} cannot assume that literals in $S_N$ do not affect the examples' classifications and thus they are not ignored and $\psi_S^{Rs}$ is used instead of $\psi_{S\setminus S_N}^{Rs}$.

\begin{algorithm}[t]
\small
\DontPrintSemicolon
\LinesNumbered

$sets$ = $\{(\{l \mid e \models l \land con \not \models l\},\emptyset)\}$\;
\For{$(Ng, Rs) \in sets$}{
    \If{sat($\psi^{Ng}_{con}$)}{
        \If{isPos $\&\&$ askUser(ex($\psi^{Ng}_{con}$), $E_P$, $E_N$) == neg}{\Return{true}}
        \ElseIf{!isPos${\&\&}$askUser(ex($\psi^{Ng}_{con}$),$E_P$,$E_N$)==pos}{\Return{true}}
        }

    $core$ = $unsatCore(\psi^{Ng}_{con})\setminus (con \cup Rs))$\;
    $sets$ = $sets \setminus \{(Ng, Rs)\}$\;
    $sets$ = $ sets \cup \{(Ng \setminus \{Q\}, Rs \cup (core \setminus \{Q\}) \mid Q \in core\}$\;
}

\Return{false}


\caption{overgen($con$, $e$, $E_P$, $E_N$, isPos)}\label{Alg:overgen}
\end{algorithm}

\subsection{The Overgen Operation}\seclabel{overgen}

The \scode{overgen} operation (\algref{overgen}) takes a conjunction $con$, the example $e$ from which $con$ was generalized, $E_P$ and $E_N$, and a flag \emph{isPos} indicating whether $e$ is a positive or a negative example. It returns \scode{true} or \scode{false} to indicate whether $con$ is an over-generalization of $e$.

\para{Main Idea}
If $con$ over-generalizes $e$, there are (at least) two sub-concepts containing $e$ and captured by conjunctions that include $con$ and additional literals. These literals are:
  \begin{inparaenum}[(i)]
    \item satisfied by $e$, and
    \item not implied by $con$.
  \end{inparaenum}
  Also, if $con$ is an over-generalization, there are negative examples satisfying it. Namely, there are examples satisfying $con$ but not any of the conjunctions of the sub-concepts. In particular, if there is an example:
  \begin{inparaenum}[(i)]
    \item satisfying $con$, and
    \item not satisfying any of the other literals satisfied by $e$ (which are not implied by $con$),
  \end{inparaenum}
  then such example must be a negative example (or a positive example if \emph{isPos} is false) because it does not satisfy any of the other literals in the sub-concepts' conjunctions. Thus, such example can be used to determine whether $con$ over-generalizes: if the user classifies it as positive (or negative if \emph{isPos} is false), $con$ does not over-generalize, and if it is negative, $con$ over-generalizes.

    However, due to the dependency, there might not be such example, in which case \scode{overgen} generates all satisfiable formulas that negate as many literals as possible (which must include a formula that does not satisfy any of the sub-concepts' conjunctions, if $con$ over-generalizes). Then, it presents the user the corresponding examples, and if one of the examples is negative (or positive if \emph{isPos} is false), it determines that $con$ over-generalizes; otherwise, it determines that $con$ does not over-generalize.

\para{Implementation} The implementation of \scode{overgen} resembles the \scode{getMinLiteralNExamples} operation only that it begins with negating all literals and diminishes this set if it is unsatisfiable (instead of beginning with one literal and extending its set). To this end, it maintains a set of tuples called $sets$ whose tuples consist of:
\begin{inparaenum}[(i)]
  \item a set $Ng$ of literals that have to be negated, and
  \item a subset of $Ng$, $Rs$, whose literals cannot be removed from $Ng$, as they are examining a certain possibility of relaxation (similarly to the $Rs$ sets in \scode{getMinLiteralNExamples}).
\end{inparaenum}
Initially, $sets$ contains a single tuple whose $Ng$ is the set of all tuples satisfied by $e$ and not implied by $con$ and $Rs$ is the empty set (there are no constraints yet on which literals cannot be removed). Then, a loop iterates the tuples in $sets$. For each tuple in $sets$, if $con$ and the literals in $Ng$ are satisfiable, i.e., $\psi_{con}^{Ng}$ (in its C-\alg\ variation) is satisfiable, an example is presented to the user, and if it is a negative example (or a positive example if \emph{isPos} is false), true is returned to indicate that $con$ over-generalizes.
If $\psi_{con}^{Ng}$ is unsatisfiable, then the tuple is replaced with a set of tuples, each considers a different possibility to relax $Ng$, that is removing a literal from the UNSAT core and obligating the other literals in the UNSAT core to remain in the relaxed $Ng$. Similarly to \scode{getMinLiteralNExamples}, the UNSAT core is cleaned from $con$ and $Rs$, from the same reason that \scode{getMinLiteralNExamples} removes $S_P$ and $Rs$ from the cores.

The loop terminates after all tuples correspond to satisfiable formulas, and their examples were classified as positive by the user (or negative if \emph{isPos} is false). In this case, it is guaranteed that $con$ does not over-generalize, and thus \scode{false} is returned.

 \subsection{Gen-\alg\ Correctness and Guarantees}
 We next state that Gen-\alg\ learns DNF formulas (more precisely, formulas that are close to DNF and can be easily transformed to DNFs), and that it learns with a minimal number of questions. Proofs are provided in~\appref{proofs}.

 \begin{theorem}\thelabel{dnfpred}
  Given $D$, $S$, $\varphi_C$ an unknown DNF specification over $S$, and initial positive and negative examples $E_P$ and $E_N$. Gen-\alg\ is a DNF-\name\ algorithm, i.e., it learns a specification $\varphi$ over $S$ such that: $\forall d. \varphi_C(d) \leftrightarrow \varphi(d)$.
\end{theorem}

\begin{theorem}\thelabel{predcomplex2}
  Given $D$, literals $S$ of size $n$, $\varphi_C$ an unknown DNF specification over $S$, and initial examples $E_P$
   and $E_N$. If Gen-\alg\ presents $\Omega(f(n))$ questions for some function $f$, any DNF-\name\ algorithm presents $\Omega(f(n))$ questions.
\end{theorem}

\section{Evaluation}\seclabel{eval}
In this section, we evaluate \alg\ on an extension of the example from~\secref{Overview} and on two new applications.

\subsection{\alg\ Guided Traversal vs. Unguided Traversals}
In this section, we experimentally demonstrate the importance of a guided traversal dependent on the vocabulary size (i.e., the predicates) instead of the domain size. To this end, we show that unguided traversals can present an exponential number of questions (in the number of predicates), even when a linear number suffices.

\para{Unguided Traversals} We consider CEGIS~\cite{solar2008program} and Oracle-Guided Synthesis~\cite{BitManipulation:2010} that perform unguided traversals:
\begin{itemize}[nosep,nolistsep]
  \item Oracle-Guided Synthesis~\cite{BitManipulation:2010} has the same interaction model as \alg, where only membership queries are permitted. It begins by finding the set of formulas consistent with the initial examples, and iteratively prunes the space until only equivalent formulas remain (i.e., the concept formula). To prune the space, it \emph{arbitrarily} selects two non-equivalent formulas consistent with the current examples, presents an example that distinguishes them, and prunes the space based on the user's output.
  \item CEGIS~\cite{solar2008program} has a different interaction model, where validation queries may also be presented, i.e., the user may be asked to confirm the specification. It begins by finding a formula consistent with the initial example, asks the user whether this is the correct concept, if so it terminates, otherwise it asks the user for an example eliminating this candidate, and repeats this process. We compare to this algorithm even though it has a different interaction model to emphasize that \emph{even if the algorithm may present more powerful questions, an unguided traversal may still result in a  exponential number of questions}.
\end{itemize}

We consider the following setting:
\begin{compactitem}
  \item Domain: $D=\{(x,y)\mid~0\leq x \leq 2000,0 \leq y \leq 2000\}$,
  \item Set of predicates: $S_0=P_x \cup P_y \cup \{x=y\}$, where $P_x=\{ 400a \leq x \leq 400b \mid  0 \leq a < b \leq 5\}$ and $P_y$ is identical with respect to $y$.
  \item Concept: $C_0=(800\leq x \leq 1200) \land (800\leq  y\leq 1200)$.
  \item Initial (positive) example: $(840, 840)$. There are $18$ predicates in $S_0$ satisfied by this example.
\end{compactitem}
To demonstrate that unguided traversals dramatically increase the number of questions as the number of predicates increases, in contrast to \alg, we consider six steps that modify this setting by extending $S_0$ and refining $C_0$.
The added predicates are of the form $(z\div a)$, which is satisfied if $a$ divides $z$. The steps are:
{\footnotesize
\begin{compactenum}
  \item $S_1=S_0 \cup \{(x \div 2),(y \div 2)\}$ and $C_1 = C_0$.
  \item $S_2 = S_1 \cup \{(x\div 3),(y\div 3)\}$ and $C_2 = C_1 \land (x\div 3)\land (y\div 3)$.
  \item $S_3 = S_2 \cup \{(x\div 4),(y\div 4)\}$ and $C_3 = C_2 \land (x\div 4)\land (y\div 4)$.
  \item $S_4 = S_3 \cup \{(x \div 5),(y \div 5)\}$ and $C_4 = C_3 \land (x\div 5)\land (y\div 5)$.
  \item $S_5 = S_4 \cup \{(x \div 6),(y\div 6)\}$ and $C_5 = C_4 \land (x\div 6)\land (y \div 6)$.
  \item $S_6 = S_5 \cup \{(x \div 7),(y \div 7)\}$ and $C_6 = C_5 \land (x\div 7)\land (y\div7)$.
\end{compactenum}
}
All new predicates are satisfied by the initial example, namely each step increases the number of satisfied predicates by two.

We ran \alg\ and the algorithms of \cite{BitManipulation:2010} and \cite{solar2008program} on this benchmark and counted the number of questions. For the unguided traversals, which are non-deterministic, we ran $10$ experiments and computed the average, maximum, and minimum number of questions. We also computed the \emph{increase factor} of two consecutive steps, which is the ratio between the increase in the number of questions and the increase in the number of predicates, that is: $Inc=(q_j - q_{j-1})/(p_j - p_{j-1})$, where $q_j$ is the number of questions at step $j$ and $p_j$ is the number of predicates satisfied by the initial example at step $j$.

\begin{table}\centering
\vspace{0.15cm}
\scriptsize
\ra{1.2}
\begin{tabular}{c|p{0.31cm}p{0.31cm}|p{0.31cm}p{0.4cm}p{0.31cm}p{0.31cm}|p{0.31cm}p{0.4cm}p{0.31cm}p{0.31cm}}\toprule
 & \multicolumn{2}{c}{\alg} & \multicolumn{4}{c}{Oracle-Guided~\cite{BitManipulation:2010}}& \multicolumn{4}{c}{CEGIS~\cite{solar2008program}}\\
p &  \#Q& Inc. & Avg. & Inc. & Max & Min & Avg. & Inc. & Max & Min\\
\midrule
18&	6&	&	11&		&12&		&6.9&		&7	&6\\
20&	8&	1&	14.2&	1.6&	17&	12&	8.1&	0.6&	9&	8\\
22&	10&	1&	21.9&	3.9&	30&	17&	13.7&	2.8&	14&	13\\
24&	12&	1&	24.7&	1.4&	27&	23&	12.6&	-0.6&	28&	10\\
26&	14&	1&	33.4&	4.4&	55&	25&	26&	6.7	&60&	12\\
28&	14&	0&	32.2&	-0.6&	51&	22&	18.1&	-4.0&	34&	12\\
30&	16&	1&	55.2&	11.5&	88&	36&	59.4&	20.7&	80&	43\\
\bottomrule
\end{tabular}\caption{\alg\ VS. Oracle-guided Synthesis~\cite{BitManipulation:2010} and CEGIS~\cite{solar2008program}.}\tablabel{comp}
\end{table}

\tabref{comp} shows the number of questions presented by the algorithms at each step and the increase factor (computed on the average number of questions).
The table shows that as $p$ increases (i.e., there are more consistent concepts to consider), \alg\ increases its number of questions \emph{linearly} in $p$ (demonstrated by the column \emph{Inc.}). Moreover, it never introduces more than $p/2$ questions. However, it is not the case for unguided traversals: 
 the number of questions presented by the algorithm of~\cite{BitManipulation:2010} increases drastically and inconsistently. This is also true for CEGIS, though at the first steps it enjoys the advantage of being able to ask validation queries.

\subsection{Technical Analysis Patterns}\seclabel{techanalysis}
Technical analysis, used for trading assets such as stocks, futures, and commodities, tries to predict future price movement based on:
\begin{inparaenum}[(i)]
  \item past price changes, often visualized in charts (functions mapping a finite set of consecutive dates to their corresponding prices), and
  \item special forms known as patterns.
\end{inparaenum}
The occurrence of a pattern in a chart is used as a predictor of future price trends. For example, the head and shoulders pattern in~\figref{HeadShoulders} predicts price decline. Patterns are mainly characterized by the relation between the price points and learning them can be viewed as learning conjunctive specification from examples (where charts serve as examples).

We employed C-\alg\ to learn patterns from charts.
The patterns are captured via conjunctive formulas over the less-than predicates, defined over the extreme points of the charts. For example, the head and shoulders pattern is defined over $7$ extreme points (marked in red rectangles in~\figref{HeadShoulders}), denoted by $p_0,...,p_6$, where $p_0$ is the price at the earliest time point, and $p_6$ is the price at the latest, and is defined as follows:
{\label{one}
 \[
 \begin{array}{l}
 \varphi_{HS}$=$\ltpred{p_0}{p_1} \land \ltpred{p_2}{p_1} \land \ltpred{p_1}{p_3} \land \ltpred{p_5}{p_3} \land \ltpred{p_4}{p_5} \land \ltpred{p_6}{p_5}
 \end{array}
 \]
 }
C-\alg\ enabled us to design a synthesizer that learns the pattern using C-\alg\ and then synthesizes a program in AFL, which  is the DSL of AmiBroker~\cite{AmiBroker}, a popular trading platform. Once the pattern has been learned, the details of the AFL synthesis are straightforward and beyond the scope of this paper.

We next formally define the problem of learning these patterns.

\begin{figure}
  \begin{center}
  \includegraphics[width=4.9cm]{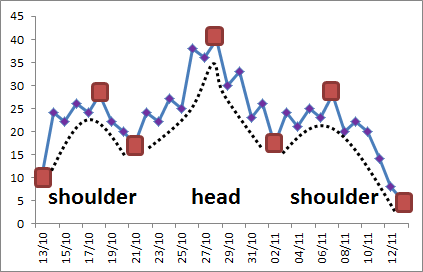}
    \caption{Head and Shoulders. }
    \figlabel{HeadShoulders}
    \end{center}
\end{figure}


\begin{definition}[\name\ in Technical Analysis Patterns]
Given a pattern $\hat{P}$ and a chart $C_{\hat{P}}$ of size $n$ following $\hat{P}$, \emph{\name\ in technical analysis patterns} is the following C-\name:
\begin{itemize}[nosep,nolistsep]
  \item The domain is charts of size $n$: $D_{\hat{P}}$=$\{(p_0,...,p_{n-1}) \mid p_i \in \mathbb{R}\}$.
  \item The predicates are $S_{\hat{P}}=\{\ltpred{p_i}{p_j}, \neg (\ltpred{p_i}{p_j}) \mid 0 \leq i, j < n\}$.
  \item The conjunctive specification $\psi_{\hat{P}}$ is satisfied by $C'$ iff $C'$ follows the pattern $\hat{P}$.
  \item $E_P=\{C_{\hat{P}}\}$, $E_N=\emptyset$.
\end{itemize}
\end{definition}

The predicates used for technical analysis satisfy Claim \ref{relativecomp} and thus C-{\alg} learns the concept with a linear number of questions.




\para{Evaluation} We evaluate C-\alg\ on common technical analysis patterns~\cite{bulkowski2012visual}.
We selected six patterns that span the space of common patterns. Though patterns are subjective and analysts define their own patterns, we believe that identifying these patterns successfully shows the effectiveness of C-\alg\ in learning patterns.


\para{Experiments} We conducted several experiments. In each we fixed a pattern $\hat{P}$, a positive example $C_{\hat{P}}$, and a goal formula $\psi$. We then let {C-\alg} learn $\psi$ interactively and counted the number of questions presented. Since patterns are subjective and definitions vary among analysts,
 for each pattern we varied the goal formulas $\psi$ over different definitions (described in~\appref{patterns}). The definitions are not inherently different, but rather differ in their strictness -- some include more constraints (predicates) than the others. In the experiments, definition $(i)$ is less strict than definition of $(j)$ for $i < j$, i.e., definition $(j)$ contains all the predicates of $(i)$ and more.

\para{Results} \tabref{learn} shows the results. The columns are: the pattern name (\emph{Pattern}),
the total number of predicates (\emph{$|S_{0}|$}); the definition used (\emph{Def.}); the number of predicates in the learned formula (\emph{$|S_{P}|$}); and the number of questions presented to the user (\emph{\#Q}). We also compare to the unguided traversal of~\cite{BitManipulation:2010} that employs the same interaction model as \alg. As this is a non-deterministic algorithm, we ran $10$ experiments and we report the average, maximum, and minimum number of questions.
The results show that C-\alg\ drastically reduces the number of questions presented, and by up to $5$ times compared to the alternative. We also measured the time taken to complete, however, since all times are in milliseconds and thus negligible, we provide them in the appendix (\appref{extendedresults}), and here provide a summary of these results. The results show that the time to generate the next question in both algorithms is always less than $700$ milliseconds, and on average around $180$ milliseconds in C-\alg\ and $100$ in the oracle-guided algorithm. C-\alg\ requires more time to generate questions, however the differences in times are not observable for human users, as the overall time is in milliseconds. The table also shows that the number of questions is correlated to the number of irrelevant predicates, namely stricter definitions require fewer questions than relaxed definitions.


\begin{table}\centering
\scriptsize
\ra{1.2}
\begin{tabular}{L{0.8cm}R{0.6cm}R{0.5cm}rrrcc}
\toprule
      &       &       &       & \textbf{CSPEX} & \multicolumn{3}{c}{\textbf{Oracle Guided}} \\
\textbf{Pattern } & \boldmath{}\textbf{ $|S_{0}|$ }\unboldmath{} & \textbf{ Def. } & \boldmath{}\textbf{ $|S_{P}|$ }\unboldmath{} & \multicolumn{1}{c}{\textbf{Num}} & \textbf{ Avg. } & \textbf{ Max } & \textbf{ Min} \\
\hline
 \multirow{5}{1.2cm}{Head and Shoulders}   & \multicolumn{1}{c}{\multirow{5}[2]{*}{42}} &  \ref{head2}  & 6     & \multicolumn{1}{c}{18} & 44.4  & 57    & 38 \\
  & \multicolumn{1}{c}{} &  \ref{head3}  & 10    & \multicolumn{1}{c}{18} & 54.8  & 61    & 43 \\
 & \multicolumn{1}{c}{} &  \ref{head4} & 10    & \multicolumn{1}{c}{17} & 61.1  & 88    & 43 \\
      & \multicolumn{1}{c}{} &  \ref{head5} & 7     & \multicolumn{1}{c}{14} & 50.6  & 71    & 40 \\
      & \multicolumn{1}{c}{} &  \ref{head1} & 6     & \multicolumn{1}{c}{12} & 58.4  & 77    & 37 \\
\hline
\multirow{5}{1.2cm}{Cup with Handle}    & \multicolumn{1}{c}{\multirow{5}[2]{*}{30}} &  \ref{cup1} & 5     & \multicolumn{1}{c}{12} & 38.3  & 48    & 27 \\
  & \multicolumn{1}{c}{} & \ref{cup2} & 6     & \multicolumn{1}{c}{12} & 43.6  & 57    & 33 \\
& \multicolumn{1}{c}{} &  \ref{cup3} & 7     & \multicolumn{1}{c}{13} & 35.1  & 41    & 28 \\
      & \multicolumn{1}{c}{} &  \ref{cup4} & 7     & \multicolumn{1}{c}{13} & 36.4  & 41    & 32 \\
      & \multicolumn{1}{c}{} &  \ref{cup5} & 5     & \multicolumn{1}{c}{10} & 38.6  & 45    & 22 \\
\hline
\multirow{4}{1.2cm}{Two Tops}  & \multicolumn{1}{c}{\multirow{4}[2]{*}{20}} &  \ref{top1} & 6     & \multicolumn{1}{c}{9} & 19.4  & 20    & 18 \\
      & \multicolumn{1}{c}{} & \ref{top2} & 6     & \multicolumn{1}{c}{9} & 18.3  & 20    & 16 \\
      & \multicolumn{1}{c}{} &  \ref{top3} & 6     & \multicolumn{1}{c}{7} & 18.7  & 19    & 17 \\
      & \multicolumn{1}{c}{} &  \ref{top4} & 4     & \multicolumn{1}{c}{6} & 17    & 17    & 17 \\
\hline
\multirow{2}{1.2cm}{Symmetrical Triangle}   & \multicolumn{1}{c}{\multirow{2}[2]{*}{42}} &  \ref{trian1} & 7     & \multicolumn{1}{c}{16} & 71.4  & 76    & 62 \\
 & \multicolumn{1}{c}{} &  \ref{trian2} & 7     & \multicolumn{1}{c}{14} & 66.5  & 78    & 37 \\
\hline
\multirow{2}{1.2cm}{Flag}  & \multicolumn{1}{c}{\multirow{2}[2]{*}{42}} & \ref{flag1} & 7     & \multicolumn{1}{c}{17} & 64    & 89    & 52 \\
      & \multicolumn{1}{c}{} &  \ref{flag2} & 6     & \multicolumn{1}{c}{16} & 54.3  & 90    & 43 \\
\hline
\multirow{2}{1.2cm}{Rectangle}    & \multicolumn{1}{c}{\multirow{2}[2]{*}{20}} &  \ref{rec1} & 6     & \multicolumn{1}{c}{8} & 27.4  & 28    & 25 \\
      & \multicolumn{1}{c}{} &  \ref{rec2} & 6     & \multicolumn{1}{c}{8} & 25.2  & 27    & 22 \\
\bottomrule
\end{tabular}%
\caption{C-\alg\ Results. Number of questions presented by C-\alg\ vs. average, maximal, and minimal number of questions presented by the Oracle-guided approach.}\tablabel{learn}
\end{table}

\subsection{Commutative Data Structure Operations}\seclabel{commspec}
We employed Gen-\alg\ for learning commutative specifications of data structures, an important task in concurrency (e.g.,~\cite{Dimitrov:2014,Herlihy:2008,Kulkarni:2011}).
A recent work~\cite{Commutatitivity:2015} shows how to learn commutative specification by
type-aware sampling over the operations' parameters and the data structure states.
The sampled parameters and data structure states are submitted to the program (which serves as an oracle) to determine whether the operations commute. Though this work was shown to be practical and correct for various data structures, there is no guarantee that all non-commutative executions are sampled, and thus the formula may be unsound. This may occur since this approach has no control over the output part of the input-output examples. To avoid such cases, their evaluation reports that at least $5000$ samples are used for every specification. As we next show, Gen-\alg\ enables to carefully select the generated examples to both guarantee that no execution is missed with a minimal number of questions (on most tested scenarios this number is less than $10$).

We next explain the task of learning commutative specifications with the example of sets.

\para{Sets} A set stores unique elements and supports the standard operations \scode{insert(e)}, \scode{remove(e)}, and \scode{contains(e)}. The operations return a flag $ret_{e}$ indicating whether they were successful: \scode{insert(e)} succeeds if $e$ was not in the set and was thus inserted, \scode{remove(e)} succeeds if $e$ was in the set and was thus removed, and \scode{contains(e)} succeeds if $e$ is in the set.

\para{Commutative Specifications} A commutative specification of two operations is a formula satisfied when the two operations commute, i.e., when the resulted set and their return values are identical regardless of the order of the operations.
For \scode{insert($e_1$)} and \scode{insert($e_2$)}, if $e_1 \neq e_2$, then the insertions do not affect each other and thus commute. If $e_1 = e_2$ then they commute only if their element was already in the set and none of them inserted. Namely, their commutative specification is:
$$\varphi_{insert(e_1),insert(e_2)}=(e_1 \neq e_2) \vee (\neg ret_{e_1} \wedge \neg ret_{e_2})$$




\para{DNF-\name\ of Commutative Specifications} \emph{\name\ of commutative specifications} gets as input two formulas capturing the operations' behaviour and defined over
\begin{inparaenum}[(i)]
  \item the data structure before the operation,
  \item the data structure after the operation,
  \item the operation's parameters, and
  \item the return value.
\end{inparaenum} For example, \scode{insert(e)} is captured by the following formula:
\[
\begin{array}{lll}
\varphi_{insert}(ds,ds',e,ret)&=&(e\in ds \implies ds' = ds \cup \{e\} \land ret) \\
&&\land (e\notin ds \implies ds' = ds \land \neg ret)
\end{array}
\]
 The data structures (i.e., $ds$, $ds'$ in the formula above) are captured using functions. For example, for $ds$ which is a set, it is defined by the function $Q:Elem \rightarrow \{0,1\},Q(e)=1 \text{ iff e is in the set}$, which can be encoded as a formula.
 We next formalize the task of learning commutative specifications formally.


\begin{definition}[Commutative Specifications \name]
\sloppy
Given two operations' formulas, $\varphi_{op_1}(ds,ds',e^1_1,...,e^1_k,ret_1)$ and $\varphi_{op_2}(ds,ds',e^2_1,...,e^2_m,ret_2)$, capturing the operation behaviours and defined over:
 \begin{inparaenum}[(i)]
   \item $ds$ and $ds'$: the state of the data structure before and after the operation (resp.),
   \item $e_1,...,e_k$: the parameters, and
 \item $ret$: the return value,
 \end{inparaenum}
  \emph{commutative specifications \name} is defined as follows:
\begin{itemize}[nosep,nolistsep]
  \item The domain is the set of feasible executions consisting of a single invocation of each operation. Each execution is captured by a tuple consisting of:
      \begin{inparaenum}[(i)]
   \item $ds$, $ds'$, $ds''$: the state of the data structure before the operations, after the first executed operation, and after the second executed operation (resp.),
   \item $e^1_1,...,e^1_k$ and $e^2_1,...,e^2_k$: the parameters of the operations, and
 \item $ret_1$ and $ret_2$: the return values of the operations.
 \end{inparaenum}
      Formally, \[
      \begin{array}{l}
      D=\{(ds,ds',ds'',e^1_1,...,e^1_k,e^2_1,...,e^2_m,ret_1,ret_2) \mid  \\~~~~~~~~~~~~~~~~~~~~~[\varphi_{op_1}(ds,ds',e^1_1,...,e^1_k,ret_1) \land \\~~~~~~~~~~~~~~~~~~~~~\varphi_{op_2}(ds',ds'',e^2_1,...,e^2_m,ret_2)] \vee \\~~~~~~~~~~~~~~~~~~~~~[\varphi_{op_2}(ds,ds',e^2_1,...,e^2_m,ret_2) \land \\~~~~~~~~~~~~~~~~~~~~~\varphi_{op_1}(ds',ds'',e^1_1,...,e^1_k,ret_1)] \}
      \end{array}\]
  \item The set of predicates $S$ contains the pairwise relative comparisons over $<,=$ of all numeric values (elements and return values) and the two states of boolean values (return values). 
  \item A feasible execution is a positive example if it is commutative, captured by satisfying the formula $\psi_{comm}$:
  \[ \begin{array}{l}
  \psi_{comm}=(\varphi_{op_1}(ds,ds'_1,e^1_1,...,e^1_k,ret_1) \land
  \\
  ~~~~~~~~~~~~~~~~~~~~~\varphi_{op_2}(ds'_1,{ds{''}},e^2_1,...,e^2_m,ret_2)) \implies \\
  ~~~~~~~~~~~~~~~~~~~~~(\varphi_{op_2}(ds,ds'_2,e^2_1,...,e^2_m,ret_2) \land \\
  ~~~~~~~~~~~~~~~~~~~~~ \varphi_{op_1}(ds'_2,{ds{''}},e^1_1,...,e^1_k,ret_1))\end{array}\]
  \item $E_P$ and $E_N$ are sets of feasible executions. We set both to be empty, and let Gen-\alg\ discover the required examples.

\end{itemize}
\end{definition}


\begin{table}\centering
\vspace{0.12cm}
\scriptsize
\ra{1.2}
\begin{tabular}{p{0.8cm}p{0.6cm}p{0.6cm}rrrrrr}\toprule
DS & Op1 & Op2 & pos & neg & Q & $|S_{0}|$ & $|S_{P}|$ & Time\\  \midrule
Set & Con & Con & 1 & 0 & 6 & 4 & 0 & 49\\
Set & Con & Add & 2 & 1 & 10 & 12 & 2 & 236\\
Set & Con & Rem & 2 & 1 & 11 & 12 & 2 & 224\\
Set & Con & Size & 1 & 0 & 5 & 3 & 0 & 31\\
Set & Add & Add & 2 & 2 & 9 & 16 & 3 & 403\\
Set & Add & Rem & 1 & 1 & 9 & 8 & 1 & 243\\
Set & Add & Size & 1 & 1 & 6 & 6 & 1 & 113\\
Set & Rem & Rem & 2 & 2 & 9 & 16 & 3 & 376\\
Set & Rem & Size & 1 & 1 & 6 & 6 & 1 & 109\\
Set & Size & Size & 1 & 0 & 1 & 2 & 0 & 8\\
\hline
Queue & Top & Top & 1 & 0 & 1 & 2 & 0 & 13\\
Queue & Top & Push & 1 & 1 & 3 & 4 & 1 & 75\\
Queue & Top & Pop & 1 & 1 & 3 & 4 & 1 & 77\\
Queue & Top & Size & 1 & 0 & 3 & 2 & 0 & 20\\
Queue & Push & Push & 1 & 1 & 3 & 4 & 1 & 153\\
Queue & Push & Pop & 0 & 1 & 3 & 2 & 0 & 118\\
Queue & Push & Size & 0 & 1 & 3 & 2 & 0 & 36\\
Queue & Pop & Pop & 1 & 1 & 3 & 4 & 1 & 130\\
Queue & Pop & Size & 1 & 1 & 3 & 4 & 1 & 68\\
Queue & Size & Size & 1 & 0 & 1 & 2 & 0 & 9\\
\hline
Reg & Get & Get & 1 & 0 & 1 & 2 & 0 & 8\\
Reg & Get & Set & 1 & 2 & 4 & 18 & 2 & 405\\
Reg & Set & Set & 1 & 3 & 20 & 48 & 3 & 6s\\
\hline
Map & Get & Get & 1 & 0 & 31 & 12 & 0 & 451\\
Map & Put & Get & 2 & 1 & 154 & 60 & 2 & 17s\\
Map & Put & Put & 2 & 3 & 842 & 150 & 4 & 9m\\
\bottomrule
\end{tabular}
\caption{Gen-\alg\ Evaluation Results.}\tablabel{learn2}
\end{table}


\para{Evaluation} We evaluated Gen-\alg\ on commutative specifications of four common data structures: set, map, queue, and max register, with their standard operations:
\begin{itemize}[nosep,nolistsep]
  \item Set: \scode{contains(k)}, \scode{add(k)}, \scode{remove(k)}, and \scode{size()}.
  \item Map: \scode{get(k)} and \scode{put(k,v)}, both return the value at position $k$ (current or former), or $0$ if $k$ was not set before.
   \item Queue: \scode{top()}, \scode{push(k)}, \scode{pop()}, and \scode{size()}, where \scode{top} does not affect the queue and \scode{push} does not return any value.
   \item Max register: \scode{get()} and \scode{set(k)} where \scode{set} updates the register only if $k$ is greater than its current value.
\end{itemize}
For each, we learned the specification of every pair of operations. 

\para{Results} \tabref{learn2} shows the results. The columns are the data structure (\emph{DS}), the operations used (\emph{Op1},\emph{Op2}),
the number of positive (pos) and negative examples (neg) \alg\ used for generalizing to conjunctions (it does not include the examples it discovered while learning in C-\alg, these examples are presented by the next column),
the number of questions ($Q$),
the total number of literals ($|S_{0}|$), the number of literals in the learned formula ({$|S_{P}|$), and the time Gen-\alg\ ran in milliseconds, unless followed by $s$ or $m$ to indicate that time is in seconds or minutes.

 \tabref{learn2} shows that Gen-\alg\ completes fast when the initial number of literals is small. Also, even when there are many dependencies, the number of questions is significantly lower compared to previous work~\cite{Commutatitivity:2015} that required for the very least $5000$ examples. Lastly,
 the specifications that have no negative behaviours (i.e., the ones that do not modify the data structure) complete after Gen-\alg\ observed a single (positive) example: for this example it learned the formula \scode{true}, determined that \scode{true} is not an over-generalization, and completed.

\section{Related Work}\seclabel{Related}
In this section we discuss work that is most closely related to ours.

\para{Learning Exact Specifications from Examples} Oracle-Guided Synthesis~\cite{BitManipulation:2010} is the closest setting to learning exact specifications. In this work, the space of programs (in our setting, specifications) is examined by iteratively searching for two programs with a \emph{distinguishing input}, asking the user for its outputs, and pruning inconsistent programs, until converging to semantically-equivalent programs. Unfortunately, this may require an exponential number of questions. The reason is that while ideally every question prunes half of the space, this occurs only when the observed examples imply the classification of some predicate. 
To address this issue, the authors of~\cite{BitManipulation:2010} suggest users to begin with a small number of components (i.e., predicates) and gradually extend it until the resulting programs (specifications) capture their intent. Unfortunately, this requires users to validate the programs, which is undesirable and contradicts the premise of our work. Another work that learns exact specification is CEGIS~\cite{solar2008program}, however it assumes that the user is expert and can read the synthesized solution, confirm if it is correct or provide an example to eliminate this solution. Even though it can enjoy more powerful questions, it may present an exponential number of questions, as the examples the user provide may lead to pruning only a few hypotheses from the hypothesis space.

\para{Program Synthesis}
The interest in program synthesis has grown dramatically over the years, and especially in the setting of synthesis from examples
(e.g.,~\cite{Gulwani:2010,Lau03,Sarma:2010,Harris:PLDI11,Gulwani:2011:ASP:1926385.1926423,Gulwani:CACM12,Singh:VLDB12,Yessenov13,Recursive:2013,Sai:2013,Aditya13,FlashExtract:14,Feser:2015,Barowy:2015,Polozov:2015,Singh:2016,Raychev:2016}). However, these works focus on synthesizing programs \emph{consistent with the provided examples}, and do not necessarily capture the user intent. Naturally, this implies that the complexity analysis of all these algorithms is dependent on the number of provided examples, which enables them to be polynomial or even linear. However, the task of guaranteeing exactness is more complex as we need to also reason about examples which \emph{were not given as input}. Such examples may trigger questions by \alg, and thus the asymptotic complexity worsens. A different line of synthesis work (i.e., constrained-based) guarantees exactness but requires the user to provide the specification (e.g.,~\cite{Sketch:PLDI08,BoardSingh:2011:SDS,alur-fmcad13,BornholtTGC16}). Unfortunately, this is known to be complex and error-prone.


\para{Relationship with Learning} Exact learning from examples (\name) is closely related to \emph{query learning}~\cite{Angluin:1988}, which learns functions over input variables. Query learning is close but not identical, since {\name} learns boolean functions over \emph{arbitrary} predicates, which to the best of our knowledge is not the setting of query learning in any of its forms (e.g., DNF over inputs, automata, polynomials).

In the context of query learning, various results have been obtained for different interaction models. In particular, there has been a lot of work on query learning with \emph{equivalence queries} (e.g.,~\cite{Beimel:2000,Abasi2014}) that ask the user to validate the formula correctness (in addition to membership tests that ask users to classify selected inputs), and are not allowed in {\name}. Works that do not use equivalence queries typically do not guarantee exact learning (e.g.,~\cite{Valiant:1984}).

When the shape of the hypothesis space is \emph{restricted to combinations of independent monomials}, classical results due to Goldman and Kearns~\cite{Goldman:1995} provide lower-bounds on the required number of questions, and define the notion of a \emph{teaching dimension}. Intuitively, \emph{``the teaching dimension of a concept class is the minimum number of examples a teacher must reveal to uniquely identify any concept in the class''}~\cite{Goldman:1995}. In this paper, we show how to obtain similar results for hypothesis spaces of formulas over first-order predicates. We generalize the results obtained for monomials~\cite{Goldman:1995} by providing algorithms that are guaranteed to ask the minimal number of questions, even when there are dependencies between predicates. The beauty of our algorithms is that they do not require the user to understand the dependencies between predicates, and instead rely on the computation of minimal unsat cores during the algorithm. This allows us to guarantee convergence with a minimal number of questions. We are not aware of prior work which can address this problem. Thus, we believe our work is a contribution to query learning as well.

Specifically, for the concept class of conjunctions over monomials, it is known (\cite[Theorem 12]{Goldman:1995}) that the teaching dimension is linear in the number of examples. This theorem, and the corresponding simple algorithm, do not work for formulas over first-order predicates due to potential correlations between predicates. The C-SPEX algorithm of \secref{class} obtains similar results for the more general case of first-order predicates (see~\theref{predcomplexlinear}). As can be seen in \secref{techanalysis}, this has immediate practical implications, as the algorithm using monomials is not applicable, and the previously known oracle-guided algorithm asks a significantly larger number of questions. The results we obtain for DNF formulas over first-order predicates (\theref{predcomplex2}) similarly generalize their results for DNF formulas over monomials. The results we obtain for DNF are of direct practical value as can be seen in \secref{commspec} where our approach learns the exact specification with significantly fewer queries compared to the previous non-exact approach.

\para{Learning Specifications} The task of learning specifications from a given program was studied using both static and dynamic techniques (e.g.,~\cite{Ernst:2007,Godefroid:2012,Gupta:2009,Sharma:2014,Sharma:2013,Garg2014,Nguyen:2014}). The setting where a program is provided is inherently different from ours.

\para{Concept Learning} {\alg} is inspired by concept learning~\cite{Mitchell82}, which is the task of learning a concept from classified examples, where concepts are drawn from a hypothesis space (known as version space). {\alg} is a novel algorithm for exact learning, generating examples such that convergence to a single hypothesis is guaranteed with a minimal number of examples.

\para{Stream Pattern Detection} Many trading software platforms provide DSLs for traders (e.g., \href{www.metaquotes.net}{MetaTrader}, \href{www.metastock.com}{MetaStock}, \href{www.ninjatrader.com}{NinjaTrader}) and further DSLs exists, e.g., CPL~\cite{CPL}, a Haskell-based high-level language designed for chart pattern queries and enabling fuzzy constraints and pattern composition. However, all require users to program, including programming (and thus understanding) the patterns' mathematical specification. Other languages support queries for streams, e.g., \emph{SASE}~\cite{Wu:2006} for RFID streams, \emph{Cayuga}~\cite{Brenna:2007} for detecting complex patterns, \emph{SPL}~\cite{Hirzel:2013}, IBM's stream processing language, StreamInsight~\cite{AFA}, Microsoft's stream processing language, and \emph{ActiveSheets}~\cite{vaziri2014stream} processing streams from within spreadsheets. 
However, all require users to mathematically express the detection condition and program the detector.

\section{Conclusion}
In this paper, we explored exact learning with a minimal number of examples for specifications over first-order predicates. Learning specifications over first-order predicates is practically important, especially for programming by examples. Learning with a minimal number of examples is important for reducing end user effort.

We show that in this setting, classical results on monomials cannot be used, due to the potential correlations between predicates. We therefore present an interactive learning algorithm {\alg} that is guaranteed to ask the user a minimal number of questions, without making a priori assumptions on the relationships between predicates. We present several variations of {\alg} that can be applied to conjunctive, disjunctive, and DNF specifications. We further show that for certain predicate classes C-\alg\ is guaranteed to ask a number of questions that is linear in the number of predicates.

We have implemented {\alg} and applied it to two different application domains: pattern detection for technical analysts and data structure properties. Experimental results show that our synthesizer  learns the exact hypothesis while presenting dramatically fewer questions than previous work.

\ignore{
We addressed the problem of learning an exact hypothesis from examples classified by a user, showing the problem to be EXPTIME. We then presented an interactive learning algorithm, {\alg}, guaranteed to ask the user a minimal number of questions. \alg\ has three variations: Gen-\alg\ (for general specifications), C-\alg\ (for conjunctive specifications), and D-\alg\ (for disjunction specifications). We further showed that for the class of conjunctive specifications over relative comparison predicates, C-\alg\ is \emph{linear} in the number of presented questions. We implemented \alg\ and applied it to two different application domains: pattern detection for technical analysts and data structure properties. Experimental results show that our synthesizer presents a small number of questions, while successfully learning the exact hypothesis.
}

\bibliographystyle{nicenat}
\bibliography{bib}

\begin{thebibliography}{54}
\providecommand{\natexlab}[1]{#1}
\providecommand{\url}[1]{\texttt{#1}}
\expandafter\ifx\csname urlstyle\endcsname\relax
  \providecommand{\doi}[1]{doi: #1}\else
  \providecommand{\doi}{doi: \begingroup \urlstyle{rm}\Url}\fi

\bibitem[Abasi et~al.(2014)Abasi, Bshouty, and Mazzawi]{Abasi2014}
H.~Abasi, N.~H. Bshouty, and H.~Mazzawi.
\newblock \emph{On Exact Learning Monotone DNF from Membership Queries}, pages
  111--124.
\newblock Springer International Publishing, Cham, 2014.
\newblock URL \url{http://dx.doi.org/10.1007/978-3-319-11662-4_9}.

\bibitem[Albarghouthi et~al.(2013)Albarghouthi, Gulwani, and
  Kincaid]{Recursive:2013}
A.~Albarghouthi, S.~Gulwani, and Z.~Kincaid.
\newblock Recursive program synthesis.
\newblock In N.~Sharygina and H.~Veith, editors, \emph{Proceedings of the 25th
  International Conference on Computer Aided Verification, CAV 2013}, pages
  934--950, 2013.
\newblock URL \url{http://dx.doi.org/10.1007/978-3-642-39799-8_67}.

\bibitem[Alur et~al.()Alur, Bodik, Juniwal, Martin, Raghothaman, Seshia, Singh,
  Solar-Lezama, Torlak, and Udupa]{alur-fmcad13}
R.~Alur, R.~Bodik, G.~Juniwal, M.~M.~K. Martin, M.~Raghothaman, S.~A. Seshia,
  R.~Singh, A.~Solar-Lezama, E.~Torlak, and A.~Udupa.
\newblock Syntax-guided synthesis.
\newblock In \emph{Proceedings of Formal Methods in Computer-Aided Design
  (FMCAD)}, pages 1--8.
\newblock URL
  \url{http://ieeexplore.ieee.org/xpl/freeabs_all.jsp?arnumber=6679385}.

\bibitem[AmiBroker()]{AmiBroker}
AmiBroker.
\newblock \url{https://www.amibroker.com/}.

\bibitem[Anand et~al.(2001)Anand, Chin, and Khoo]{CPL}
S.~Anand, W.-N. Chin, and S.-C. Khoo.
\newblock Charting patterns on price history.
\newblock In \emph{Proceedings of the Sixth ACM SIGPLAN International
  Conference on Functional Programming}, ICFP '01, pages 134--145, 2001.
\newblock URL \url{http://doi.acm.org/10.1145/507635.507653}.

\bibitem[Angluin(1988)]{Angluin:1988}
D.~Angluin.
\newblock Queries and concept learning.
\newblock \emph{Machine Learning}, 2\penalty0 (4):\penalty0 319--342, 1988.
\newblock URL \url{http://dx.doi.org/10.1007/BF00116828}.

\bibitem[Angluin et~al.(1993)Angluin, Hellerstein, and Karpinski]{Angluin:1993}
D.~Angluin, L.~Hellerstein, and M.~Karpinski.
\newblock Learning read-once formulas with queries.
\newblock \emph{J. ACM}, 40\penalty0 (1):\penalty0 185--210, Jan. 1993.
\newblock URL \url{http://doi.acm.org/10.1145/138027.138061}.

\bibitem[Barowy et~al.(2015)Barowy, Gulwani, Hart, and Zorn]{Barowy:2015}
D.~W. Barowy, S.~Gulwani, T.~Hart, and B.~Zorn.
\newblock Flashrelate: Extracting relational data from semi-structured
  spreadsheets using examples.
\newblock In \emph{Proceedings of the 36th ACM SIGPLAN Conference on
  Programming Language Design and Implementation}, PLDI '15, pages 218--228,
  2015.
\newblock URL \url{http://doi.acm.org/10.1145/2737924.2737952}.

\bibitem[Beimel et~al.(2000)Beimel, Bergadano, Bshouty, Kushilevitz, and
  Varricchio]{Beimel:2000}
A.~Beimel, F.~Bergadano, N.~H. Bshouty, E.~Kushilevitz, and S.~Varricchio.
\newblock Learning functions represented as multiplicity automata.
\newblock \emph{J. ACM}, 47\penalty0 (3):\penalty0 506--530, May 2000.
\newblock URL \url{http://doi.acm.org/10.1145/337244.337257}.

\bibitem[Bornholt et~al.(2016)Bornholt, Torlak, Grossman, and
  Ceze]{BornholtTGC16}
J.~Bornholt, E.~Torlak, D.~Grossman, and L.~Ceze.
\newblock Optimizing synthesis with metasketches.
\newblock In \emph{Proceedings of the 43rd Annual ACM SIGPLAN-SIGACT Symposium
  on Principles of Programming Languages}, POPL '16, pages 775--788, 2016.
\newblock URL \url{http://doi.acm.org/10.1145/2837614.2837666}.

\bibitem[Brenna et~al.(2007)Brenna, Demers, Gehrke, Hong, Ossher, Panda,
  Riedewald, Thatte, and White]{Brenna:2007}
L.~Brenna, A.~Demers, J.~Gehrke, M.~Hong, J.~Ossher, B.~Panda, M.~Riedewald,
  M.~Thatte, and W.~White.
\newblock Cayuga: A high-performance event processing engine.
\newblock In \emph{Proceedings of the 2007 ACM SIGMOD International Conference
  on Management of Data}, SIGMOD '07, pages 1100--1102, 2007.
\newblock URL \url{http://doi.acm.org/10.1145/1247480.1247620}.

\bibitem[Bulkowski(2012)]{bulkowski2012visual}
T.~Bulkowski.
\newblock \emph{Visual Guide to Chart Patterns}.
\newblock Bloomberg Financial. 2012.

\bibitem[Bulkowski(2005)]{EncPatterns}
T.~N. Bulkowski.
\newblock \emph{Encyclopedia of Chart Patterns}.
\newblock Wiley, 2nd edition, 2005.

\bibitem[Chandramouli et~al.()Chandramouli, Goldstein, and Maier]{AFA}
B.~Chandramouli, J.~Goldstein, and D.~Maier.
\newblock High-performance dynamic pattern matching over disordered streams.
\newblock In \emph{VLDB '10}.

\bibitem[Cimatti et~al.(2011)Cimatti, Griggio, and Sebastiani]{CimattiGS11}
A.~Cimatti, A.~Griggio, and R.~Sebastiani.
\newblock Computing small unsatisfiable cores in satisfiability modulo
  theories.
\newblock \emph{J. Artif. Intell. Res. {(JAIR)}}, 40, 2011.

\bibitem[Das~Sarma et~al.(2010)Das~Sarma, Parameswaran, Garcia-Molina, and
  Widom]{Sarma:2010}
A.~Das~Sarma, A.~Parameswaran, H.~Garcia-Molina, and J.~Widom.
\newblock Synthesizing view definitions from data.
\newblock In \emph{Proceedings of the 13th International Conference on Database
  Theory}, ICDT '10, pages 89--103, 2010.
\newblock URL \url{http://doi.acm.org/10.1145/1804669.1804683}.

\bibitem[De~Moura and Bj{\o}rner(2008)]{DeMoura:2008}
L.~De~Moura and N.~Bj{\o}rner.
\newblock Z3: An efficient {SMT} solver.
\newblock In \emph{Proceedings of the 14th International Conference on Tools
  and Algorithms for the Construction and Analysis of Systems},
  TACAS'08/ETAPS'08, pages 337--340. 2008.
\newblock URL \url{http://dl.acm.org/citation.cfm?id=1792734.1792766}.

\bibitem[Dimitrov et~al.(2014)Dimitrov, Raychev, Vechev, and
  Koskinen]{Dimitrov:2014}
D.~Dimitrov, V.~Raychev, M.~Vechev, and E.~Koskinen.
\newblock Commutativity race detection.
\newblock In \emph{Proceedings of the 35th ACM SIGPLAN Conference on
  Programming Language Design and Implementation}, PLDI '14, pages 305--315,
  2014.
\newblock URL \url{http://doi.acm.org/10.1145/2594291.2594322}.

\bibitem[Ernst et~al.(December, 2007)Ernst, Perkins, Guo, McCamant, Pacheco,
  Tschantz, and Xiao]{Ernst:2007}
M.~D. Ernst, J.~H. Perkins, P.~J. Guo, S.~McCamant, C.~Pacheco, M.~S. Tschantz,
  and C.~Xiao.
\newblock The daikon system for dynamic detection of likely invariants.
\newblock \emph{Sci. Comput. Program.}, December, 2007.

\bibitem[Feser et~al.(2015)Feser, Chaudhuri, and Dillig]{Feser:2015}
J.~K. Feser, S.~Chaudhuri, and I.~Dillig.
\newblock Synthesizing data structure transformations from input-output
  examples.
\newblock In \emph{Proceedings of the 36th ACM SIGPLAN Conference on
  Programming Language Design and Implementation}, PLDI '15, pages 229--239,
  2015.
\newblock URL \url{http://doi.acm.org/10.1145/2737924.2737977}.

\bibitem[Frankle et~al.(2016)Frankle, Osera, Walker, and
  Zdancewic]{Frankle:2016}
J.~Frankle, P.-M. Osera, D.~Walker, and S.~Zdancewic.
\newblock Example-directed synthesis: A type-theoretic interpretation.
\newblock In \emph{Proceedings of the 43rd Annual ACM SIGPLAN-SIGACT Symposium
  on Principles of Programming Languages}, POPL '16, pages 802--815, 2016.
\newblock URL \url{http://doi.acm.org/10.1145/2837614.2837629}.

\bibitem[Garg et~al.(2014)Garg, L{\"o}ding, Madhusudan, and Neider]{Garg2014}
P.~Garg, C.~L{\"o}ding, P.~Madhusudan, and D.~Neider.
\newblock \emph{{ICE}: A Robust Framework for Learning Invariants}, pages
  69--87.
\newblock Springer, 2014.
\newblock URL \url{http://dx.doi.org/10.1007/978-3-319-08867-9_5}.

\bibitem[Gehr et~al.(2015)Gehr, Dimitrov, and Vechev]{Commutatitivity:2015}
T.~Gehr, D.~Dimitrov, and M.~Vechev.
\newblock Learning commutativity specifications.
\newblock In D.~Kroening and S.~C. P{\u{a}}s{\u{a}}reanu, editors,
  \emph{Proceedings of the 27th International Conference on Computer Aided
  Verification, CAV 2015}, pages 307--323. 2015.
\newblock URL \url{http://dx.doi.org/10.1007/978-3-319-21690-4_18}.

\bibitem[Godefroid and Taly(2012)]{Godefroid:2012}
P.~Godefroid and A.~Taly.
\newblock Automated synthesis of symbolic instruction encodings from i/o
  samples.
\newblock In \emph{Proceedings of the 33rd ACM SIGPLAN Conference on
  Programming Language Design and Implementation}, PLDI '12, pages 441--452,
  2012.
\newblock URL \url{http://doi.acm.org/10.1145/2254064.2254116}.

\bibitem[Goldman and Kearns(1995)]{Goldman:1995}
S.~Goldman and M.~Kearns.
\newblock On the complexity of teaching.
\newblock \emph{J. Comput. Syst. Sci.}, 50\penalty0 (1):\penalty0 20--31, Feb.
  1995.
\newblock URL \url{http://dx.doi.org/10.1006/jcss.1995.1003}.

\bibitem[Gulwani(2010)]{Gulwani:2010}
S.~Gulwani.
\newblock Dimensions in program synthesis.
\newblock In \emph{Proceedings of the 12th International ACM SIGPLAN Symposium
  on Principles and Practice of Declarative Programming}, PPDP '10, pages
  13--24, 2010.
\newblock URL \url{http://doi.acm.org/10.1145/1836089.1836091}.

\bibitem[Gulwani(2011)]{Gulwani:2011:ASP:1926385.1926423}
S.~Gulwani.
\newblock Automating string processing in spreadsheets using input-output
  examples.
\newblock In \emph{Proceedings of the 38th Annual ACM SIGPLAN-SIGACT Symposium
  on Principles of Programming Languages}, POPL '11, pages 317--330, 2011.
\newblock URL \url{http://doi.acm.org/10.1145/1926385.1926423}.

\bibitem[Gulwani et~al.(2012)Gulwani, Harris, and Singh]{Gulwani:CACM12}
S.~Gulwani, W.~R. Harris, and R.~Singh.
\newblock Spreadsheet data manipulation using examples.
\newblock \emph{Commun. ACM}, 55\penalty0 (8):\penalty0 97--105, Aug. 2012.
\newblock URL \url{http://doi.acm.org/10.1145/2240236.2240260}.

\bibitem[Gupta et~al.(2009)Gupta, Majumdar, and Rybalchenko]{Gupta:2009}
A.~Gupta, R.~Majumdar, and A.~Rybalchenko.
\newblock From tests to proofs.
\newblock In \emph{Proceedings of the 15th International Conference on Tools
  and Algorithms for the Construction and Analysis of Systems}, TACAS '09,
  pages 262--276, 2009.
\newblock URL \url{http://dx.doi.org/10.1007/978-3-642-00768-2_24}.

\bibitem[Harris and Gulwani(2011)]{Harris:PLDI11}
W.~R. Harris and S.~Gulwani.
\newblock Spreadsheet table transformations from examples.
\newblock In \emph{Proceedings of the 32Nd ACM SIGPLAN Conference on
  Programming Language Design and Implementation}, PLDI '11, pages 317--328,
  2011.
\newblock URL \url{http://doi.acm.org/10.1145/1993498.1993536}.

\bibitem[Herlihy and Koskinen(2008)]{Herlihy:2008}
M.~Herlihy and E.~Koskinen.
\newblock Transactional boosting: A methodology for highly-concurrent
  transactional objects.
\newblock In \emph{Proceedings of the 13th ACM SIGPLAN Symposium on Principles
  and Practice of Parallel Programming}, PPoPP '08, pages 207--216, 2008.
\newblock URL \url{http://doi.acm.org/10.1145/1345206.1345237}.

\bibitem[Hirzel et~al.(2013)Hirzel, Andrade, Gedik, Jacques-Silva, Khandekar,
  Kumar, Mendell, Nasgaard, Schneider, Soul{\'e}, and Wu]{Hirzel:2013}
M.~Hirzel, H.~Andrade, B.~Gedik, G.~Jacques-Silva, R.~Khandekar, V.~Kumar,
  M.~Mendell, H.~Nasgaard, S.~Schneider, R.~Soul{\'e}, and K.-L. Wu.
\newblock {IBM} streams processing language: Analyzing big data in motion.
\newblock \emph{{IBM} J. Res. Dev.}, 57\penalty0 (3-4), 2013.

\bibitem[Investopedia()]{Investopedia}
Investopedia.
\newblock
  \url{http://www.investopedia.com/university/technical/techanalysis8.asp}.

\bibitem[Jha et~al.(2010)Jha, Gulwani, Seshia, and
  Tiwari]{BitManipulation:2010}
S.~Jha, S.~Gulwani, S.~A. Seshia, and A.~Tiwari.
\newblock Oracle-guided component-based program synthesis.
\newblock In \emph{Proceedings of the 32Nd ACM/IEEE International Conference on
  Software Engineering - Volume 1}, ICSE '10, pages 215--224, 2010.
\newblock URL \url{http://doi.acm.org/10.1145/1806799.1806833}.

\bibitem[Kulkarni et~al.()Kulkarni, Nguyen, Prountzos, Sui, and
  Pingali]{Kulkarni:2011}
M.~Kulkarni, D.~Nguyen, D.~Prountzos, X.~Sui, and K.~Pingali.
\newblock Exploiting the commutativity lattice.
\newblock In \emph{PLDI '11}.

\bibitem[Lau et~al.(2003)Lau, Wolfman, Domingos, and Weld]{Lau03}
T.~Lau, S.~A. Wolfman, P.~Domingos, and D.~S. Weld.
\newblock Programming by demonstration using version space algebra.
\newblock \emph{Mach. Learn.}, 53\penalty0 (1-2):\penalty0 111--156, Oct. 2003.
\newblock URL \url{http://dx.doi.org/10.1023/A:1025671410623}.

\bibitem[Le and Gulwani(2014)]{FlashExtract:14}
V.~Le and S.~Gulwani.
\newblock Flashextract: A framework for data extraction by examples.
\newblock In \emph{Proceedings of the 35th ACM SIGPLAN Conference on
  Programming Language Design and Implementation}, PLDI '14, pages 542--553,
  2014.
\newblock URL \url{http://doi.acm.org/10.1145/2594291.2594333}.

\bibitem[Menon et~al.()Menon, Tamuz, Gulwani, Lampson, and Kalai]{Aditya13}
A.~K. Menon, O.~Tamuz, S.~Gulwani, B.~W. Lampson, and A.~Kalai.
\newblock A machine learning framework for programming by example.
\newblock In \emph{ICML '13}.

\bibitem[Mitchell(1982)]{Mitchell82}
T.~M. Mitchell.
\newblock Generalization as search.
\newblock \emph{Artificial Intelligence}, 18\penalty0 (2):\penalty0 203 -- 226,
  1982.
\newblock URL
  \url{http://www.sciencedirect.com/science/article/pii/0004370282900406}.

\bibitem[Nguyen et~al.(2014)Nguyen, Kapur, Weimer, and Forrest]{Nguyen:2014}
T.~Nguyen, D.~Kapur, W.~Weimer, and S.~Forrest.
\newblock Using dynamic analysis to generate disjunctive invariants.
\newblock In \emph{Proceedings of the 36th International Conference on Software
  Engineering}, ICSE 2014, pages 608--619, 2014.
\newblock URL \url{http://doi.acm.org/10.1145/2568225.2568275}.

\bibitem[Polozov and Gulwani(2015)]{Polozov:2015}
O.~Polozov and S.~Gulwani.
\newblock Flashmeta: A framework for inductive program synthesis.
\newblock In \emph{Proceedings of the 2015 ACM SIGPLAN International Conference
  on Object-Oriented Programming, Systems, Languages, and Applications}, OOPSLA
  2015, pages 107--126, 2015.
\newblock URL \url{http://doi.acm.org/10.1145/2814270.2814310}.

\bibitem[Raychev et~al.(2016)Raychev, Bielik, Vechev, and Krause]{Raychev:2016}
V.~Raychev, P.~Bielik, M.~Vechev, and A.~Krause.
\newblock Learning programs from noisy data.
\newblock In \emph{Proceedings of the 43rd Annual ACM SIGPLAN-SIGACT Symposium
  on Principles of Programming Languages}, POPL '16, pages 761--774, 2016.
\newblock URL \url{http://doi.acm.org/10.1145/2837614.2837671}.

\bibitem[Sharma and Aiken(2014)]{Sharma:2014}
R.~Sharma and A.~Aiken.
\newblock From invariant checking to invariant inference using randomized
  search.
\newblock In \emph{Proceedings of the 16th International Conference on Computer
  Aided Verification - Volume 8559}, pages 88--105. 2014.
\newblock URL \url{http://dx.doi.org/10.1007/978-3-319-08867-9_6}.

\bibitem[Sharma et~al.(2013)Sharma, Gupta, Hariharan, Aiken, and
  Nori]{Sharma:2013}
R.~Sharma, S.~Gupta, B.~Hariharan, A.~Aiken, and A.~V. Nori.
\newblock \emph{Verification as Learning Geometric Concepts}, pages 388--411.
\newblock Springer Berlin Heidelberg, 2013.
\newblock URL \url{http://dx.doi.org/10.1007/978-3-642-38856-9_21}.

\bibitem[Singh and Gulwani(2012)]{Singh:VLDB12}
R.~Singh and S.~Gulwani.
\newblock Learning semantic string transformations from examples.
\newblock \emph{Proc. VLDB Endow.}, 5\penalty0 (8):\penalty0 740--751, Apr.
  2012.
\newblock URL \url{http://dx.doi.org/10.14778/2212351.2212356}.

\bibitem[Singh and Gulwani(2016)]{Singh:2016}
R.~Singh and S.~Gulwani.
\newblock Transforming spreadsheet data types using examples.
\newblock In \emph{Proceedings of the 43rd Annual ACM SIGPLAN-SIGACT Symposium
  on Principles of Programming Languages}, POPL '16, pages 343--356, 2016.
\newblock URL \url{http://doi.acm.org/10.1145/2837614.2837668}.

\bibitem[Singh and Solar-Lezama(2011)]{BoardSingh:2011:SDS}
R.~Singh and A.~Solar-Lezama.
\newblock Synthesizing data structure manipulations from storyboards.
\newblock In \emph{Proceedings of the 19th ACM SIGSOFT Symposium and the 13th
  European Conference on Foundations of Software Engineering}, ESEC/FSE '11,
  pages 289--299, 2011.
\newblock URL \url{http://doi.acm.org/10.1145/2025113.2025153}.

\bibitem[Solar-Lezama(2008)]{solar2008program}
A.~Solar-Lezama.
\newblock \emph{Program synthesis by sketching}.
\newblock ProQuest, 2008.

\bibitem[Solar-Lezama et~al.(2008)Solar-Lezama, Jones, and
  Bodik]{Sketch:PLDI08}
A.~Solar-Lezama, C.~G. Jones, and R.~Bodik.
\newblock Sketching concurrent data structures.
\newblock In \emph{Proceedings of the 29th ACM SIGPLAN Conference on
  Programming Language Design and Implementation}, PLDI '08, pages 136--148,
  2008.
\newblock URL \url{http://doi.acm.org/10.1145/1375581.1375599}.

\bibitem[Valiant(1984)]{Valiant:1984}
L.~G. Valiant.
\newblock A theory of the learnable.
\newblock \emph{Commun. ACM}, 27\penalty0 (11):\penalty0 1134--1142, Nov. 1984.
\newblock URL \url{http://doi.acm.org/10.1145/1968.1972}.

\bibitem[Vaziri et~al.(2014)Vaziri, Tardieu, Rabbah, Suter, and
  Hirzel]{vaziri2014stream}
M.~Vaziri, O.~Tardieu, R.~Rabbah, P.~Suter, and M.~Hirzel.
\newblock \emph{Stream Processing with a Spreadsheet}, pages 360--384.
\newblock 2014.
\newblock URL \url{http://dx.doi.org/10.1007/978-3-662-44202-9_15}.

\bibitem[Wu et~al.(2006)Wu, Diao, and Rizvi]{Wu:2006}
E.~Wu, Y.~Diao, and S.~Rizvi.
\newblock High-performance complex event processing over streams.
\newblock In \emph{Proceedings of the 2006 ACM SIGMOD International Conference
  on Management of Data}, SIGMOD '06, pages 407--418, 2006.
\newblock URL \url{http://doi.acm.org/10.1145/1142473.1142520}.

\bibitem[Yessenov et~al.(2013)Yessenov, Tulsiani, Menon, Miller, Gulwani,
  Lampson, and Kalai]{Yessenov13}
K.~Yessenov, S.~Tulsiani, A.~Menon, R.~C. Miller, S.~Gulwani, B.~Lampson, and
  A.~Kalai.
\newblock A colorful approach to text processing by example.
\newblock In \emph{Proceedings of the 26th Annual ACM Symposium on User
  Interface Software and Technology}, UIST '13, pages 495--504, 2013.
\newblock URL \url{http://doi.acm.org/10.1145/2501988.2502040}.

\bibitem[Zhang and Sun(2013)]{Sai:2013}
S.~Zhang and Y.~Sun.
\newblock Automatically synthesizing sql queries from input-output examples.
\newblock In \emph{Automated Software Engineering (ASE), 2013 IEEE/ACM 28th
  International Conference on}, pages 224--234, Nov 2013.

\end{thebibliography}
\appendix
\section{Proofs}\applabel{proofs}
\subsection{Section 3}
\para{Claim \ref{ICSIPE-lowerbound} Proof} Let A be an C\name\ algorithm, $D=\{(x_0,...,x_k)\mid \forall i. x_i \in \{0, 1\}\}$, $S=\{(x_0 \vee x_j), (x_0 \vee \neg x_j) \mid 1\leq j \leq k\}$, and the target formula $\varphi_C = \bigwedge_{1\leq i \leq k}(x_0 \vee l_i)$ where $l_i \in \{x_i,\neg x_i\}$ are defined later.
 Assume $E_P=E_N = \emptyset$. Examples in which $x_0=1$ are positive and do not eliminate any predicate from $S$, thus assume $A$ does not present such examples (this only helps $A$ to avoid uninformative questions).
 We prove that there is a selection of $l_i$ for which the first $2^{k-1} - 1$ examples, $d_0,...,d_{2^{n-1} - 1}$, are all negative and neither enables to infer which predicates are in $\varphi_C$.

 Base: let $d_0$ be an example (in which $x_0=0$). 
 Assume $A$ infers some classification:
 \begin{itemize}[nosep,nolistsep]
   \item If $A$ infers $x_0 \vee x_i$ is in $\varphi_C$ (for some $i$), then we set $l_i=\neg x_i$ and for some $j\neq i$ we set $l_j = x_j$ if $d_0 \models \neg x_j$ or $l_j = \neg x_j$, otherwise. Namely, $A$ inferred incorrectly. 
   \item If $A$ infers $x_0 \vee \neg x_i$ is in $\varphi_C$, we set $l_i = x_i$ and get contradiction similarly.
   \item If $A$ infers that $x_0 \vee x_i$ is not in $\varphi_C$, we set $l_i = x_i$ and get contradiction similarly.
   \item If $A$ infers that $x_0 \vee \neg x_i$ is not in $\varphi_C$, we set $l_i = \neg x_i$ and get contradiction similarly.
 \end{itemize}

  Step: Assume that the $d_0,...,d_{m-1}$ ($m\leq 2^{k-1}-1$) are classified as negative and no predicate was classified as belong or not belong to $\varphi_C$. Assume $A$ infers some classification using the example $d_{m}$ (in which $x_0=0$):

 \begin{itemize}[nosep,nolistsep]
   \item If $A$ infers $x_0 \vee x_i$ is in $\varphi_C$ (for some $i$):
    There are $2^{k-1}$ examples in which $x_i$ is $1$, however there are $2^{k-1}+1$ unclassified examples (in which $x_0=0$) since $m\leq 2^{k-1} - 1$. Thus, there exists an example $d$ in which $x_0=x_i=0$, such that $d\neq d_j$ for all $0\leq j \leq m$.
     We set $l_i = x_i$ if $d \models x_i$ and $l_i = \neg x_i$, otherwise.
    For every $0\leq j \leq m$, $d_j$ is indeed negative, i.e., $d_j \not \models \varphi_C$, since $d\neq d_j$ and thus for some $i'$ $x_{i'}^d \neq x_{i'}^{d_j}$. However, $A$ inferred incorrectly because $l_i = \neg x_i$. 
   \item If $A$ infers $x_0 \vee \neg x_i\in\varphi_C$, there is $d$ in which $x_i = 1$ and we contradict A similarly.
   \item If $A$ infers $x_0 \vee x_i \notin \varphi_C$, there is $d$ in which $x_i = 1$ and we contradict A similarly.
   \item If $A$ infers $x_0 \vee \neg x_i \notin \varphi_C$, there is $d$ in which $x_i = 0$ and we contradict A similarly.
 \end{itemize}
 \subsection{Section 4}
 \para{\theref{conpred} Proof}
 We prove in induction that during the execution of C-\alg, $\bigwedge_{l \in S \setminus S_N} l \models \varphi_C \models \bigwedge_{l \in S_P} l$ and since at the end of the execution $S\setminus S_N=S_P$ and $\varphi =  \bigwedge_{l \in S \setminus S_N} l$, it follows that $\varphi \equiv \varphi_C$.
 
 Base: initially, $S_{P}=S_{N}=\emptyset$, and thus we show $S \models \varphi_C \models \emptyset$. $S$ contains all predicates satisfied by all examples in $E_P$ and since $\varphi_C$ contains literals from the initial literals, if we assume that $S\not \models \varphi_C$, then there is a literal $l$ in $\varphi_C$ not in $S$. This means that there is $e\in E_P$ not satisfying $l$, and thus not satisfying $\varphi_C$, in contradiction to the fact that $e\in E_P$.

Step:
We assume $S\setminus S_{N} \models \varphi_C \models S_{P}$ and show that updates to $S_{P}$ or $S_{N}$ preserve these implications. Let $l$ be a literal classified. We split to cases:
\begin{itemize}[nosep,nolistsep]
  \item If $l$ is classified to $S_N$ then it must because an example $e$ was generated in which $l$ is negated and $e$ was classified positive. In this case, we show $S\setminus (S_N \cup \{l\}) \models \varphi_C$ (as $S_P$ does not change, and from the induction hypothesis it continues to hold $\varphi_C \models S_P$). To show $S\setminus (S_N \cup \{l\}) \models \varphi_C$, it suffices to show that $l \notin \varphi_C$. Assume in contradiction $l \in \varphi_C$, then since $e$ is positive, $e\models \varphi_C$, however $e\models \neg l$ -- contradiction.
  \item If $l$ is classified to $S_P$ and it was classified following another literal $l'$ was added to $S_P$ and $S_P \cup \{l'\} \models l$, then $S_P \equiv S_P \cup \{l\}$ and thus $\varphi_C \models S_P \cup \{l\}$.
  \item If $l$ is classified to $S_P$ following a set of examples all classified negative. We show $l \in \varphi_C$ and thus conclude $\varphi_C \models S_P \cup \{l\}$. Assume in contradiction $l \notin \varphi_C$. Every example was classified negative, and thus for each example $e$ either $l$ is in $\varphi_C$ or one of the other literals of $Rs_e$ is in $\varphi_C$ (otherwise the example should have been classified as positive, since the rest of the predicates of $\varphi_C$ are in $S\setminus S_N$, since $S\setminus S_N \models \varphi_C$). Assume that when examining $l$ and extracting the $i^{th}$ UNSAT core, the result was $\{\neg l, l^i_1, ..., l^i_k\}$ (note that $\neg l$ must be in the UNSAT core, since $S$ is satisfiable and $l_1,...l_k\in S$), namely $l_1^i \wedge ... \wedge l_k^i \models l$. If $l_1^i,...,l_k^i \in \varphi_C$, then $\varphi_C \models l$ which is equivalent to saying that $l$ is in $\varphi_C$, in contradiction to our assumption. Thus, in each iteration of extracting the UNSAT core, one of the literals is not in $\varphi_C$, denote it by $Q^i$. However, in this case the example satisfying $\bigwedge_i \neg Q_i \wedge \neg l \wedge \bigwedge_{l'\in S\setminus (S_N \cup \{l, Q_1,...,Q_m\})}l'$ must be a positive example since it satisfies all the literals in $\varphi_C$. However, this example was classified as negative, and thus our assumption is contradicted again and $l$ must be in $\varphi_C$.
\end{itemize}

\para{\theref{cnfpred} Proof} Dual to the previous proof.

\para{\theref{predcomplex} Proof}
We prove the theorem for the C-\alg, the proof for D-\alg\ is similar.
Let A be an C\name\ algorithm, $D$ a domain, $S$ predicates over $D$ of size $n$, an unknown target formula $\varphi_C$, $E_P$ and $E_N$ sets of positive and negative examples, and a literal $l\in S$.
We prove the following claim:
\begin{claim}
To determine $l$'s classification (i.e., in or not in $\varphi_C$) at some moment during the execution, where:
 \begin{itemize}[nosep,nolistsep]
   \item $S_P$ contains all literals already known to be part of $\varphi_C$, i.e., for all $Q \in S_P$, all positive examples satisfy $Q$.
 \item $S_N$ contains all literals already known not to be in $\varphi_C$, i.e., for all $Q \in S_N$, there exist a positive example $e'$ not satisfying it (equivalently, all positive examples satisfy every predicate in $S\setminus S_N$).
 \end{itemize}
 A must either have:
 \begin{itemize}[nosep,nolistsep]
   \item A (new) positive example $e_l$ not satisfying $l$, in which case $l$ is classified to $S_N$ (as $l$ is not in $\varphi_C$ if $e_l$ is positive), or
   \item All (new) negative examples, $e^l_1,...,e^l_k$, satisfying:
\begin{itemize}[nosep,nolistsep]
  \item $e^l_1,...,e^l_k \models S_P$,
  \item for all $1\leq i \leq k$, $e_i^l \not \models l$,
  \item for all $1\leq i \leq k$, $S_{e^l_i}$ is minimal, where $S_{e^l_i}=\{Q\in S \setminus S_N \mid e^l_i \not \models Q\}$ (i.e., every subset $S'$ of $S_{e^l_i}$ does not have an example satisfying $\neg l$ and the predicates in $S\setminus S'$), and
  \item for all $1\leq i < j \leq k$, $S_{e^l_i}\neq S_{e^l_j}$,
\end{itemize}
in which case $l$ is classified to $S_P$.
 \end{itemize}
\end{claim}
Using this claim, the theorem is proven as follows:
\alg\ presents for every $l$ the examples $e^l_1,...,e^l_k$ as described in the claim (up to the property that they are negative, which is known only after the user classifies them), and may only stop if one of them is classified as positive. Namely, these examples consist the superset of the examples \alg\ presents to classify $l$. 
Thus, and since every literal in $S$ is potentially in $\varphi_C$ (more precisely, $S_{0}$ of Line \slnref{prediter:sinit} in \algref{prediter}), this claim implies that A requires at least as many examples as \alg, namely \alg\ presents a minimal number of questions. Note that if an example $e$ is required for the classification of two literals $l_1$ and $l_2$, then it is counted only once since if $e$ was classified as positive, both $l_1$ and $l_2$ are classified as not belonging to $\varphi_C$ (in which case $e$ is not presented again), and otherwise $e$ is stored in $E_N$ and will not be presented to the user again.
We note that if $l$ is not in $\varphi_C$, \alg\ presents in the worst case all examples (if all but $e^l_k$ are negative examples), and thus it may be that occasionally A ``gets lucky'' and asks fewer questions (e.g., if its first question is $e^l_k$, A need not present $e^l_1,...,e^l_{k-1}$), however, up to the order of $e^l_1,...,e^l_k$, A asks as many questions (examples) as \alg.

Proof of the Claim.
Let $S_P$ and $S_N$ be the sets as described in the claim during some point of the execution, and let $e_1,...,e_t$ be the positive examples observed and $e'_1,...,e'_{t'}$ be the negative example observed.
Assume that A classifies $l$.
 We split to cases:
 \begin{itemize}[nosep,nolistsep]
   \item if $l$ is classified to $S_N$: we show that A must have at least one positive example not satisfying $l$. Suppose otherwise, then we show that there is a specification $\varphi_C$ containing $l$ which is consistent with the previous examples' classifications and the previous predicates' classifications, and this contradicts A's classification of $l$. Assume that A does not have a positive example not satisfying $l$, namely all other positive examples $e_1,...,e_t$ satisfy $l$. We set $\varphi_C = \bigwedge_{Q \in S : e_1 \models Q \land ... \land e_t \models Q} Q$. $\varphi_C$ is consistent with the examples:
        \begin{itemize}[nosep,nolistsep]
          \item By construction all positive examples satisfy $\varphi_C$.
         \item Every negative example $e'$ does not satisfy $\varphi_C$: suppose otherwise and suppose that there is another specification $\varphi_C'$ consistent with the classification of all the observed examples.
         For every two positive examples, literals that are satisfied by one but not by the other cannot be in $\varphi_C'$ (because otherwise one of them will not be a positive example). Thus, $e'$ cannot be a negative example because of one of the literals satisfied by one of the positive examples but not by another positive example, and there must be a predicate in $\varphi_C'$ satisfied by all positive examples but not by $e'$, however in this case this predicate is also in $\varphi_C$, i.e., $e' \not \models \varphi_C$. 
        \end{itemize}
        $\varphi_C$ is consistent with the literals:
        \begin{itemize}[nosep,nolistsep]
        \item From $S_P$ definition, every positive example satisfies all literals in $S_P$, and thus every such literal in $S_P$ is also in $\varphi_C$, as required.
        \item Every literal in $S_N$ has a positive example not satisfying it, and thus by construction it is not in $\varphi_C$, as required.
        \end{itemize}
        However, $l$ is in $\varphi_C$ (all positive examples satisfy it), namely A classified $l$ incorrectly.

   \item if $l$ is classified to $S_P$: we show that A must have all $k$ examples described in the claim to classify $l$ to $S_P$. Assume that A has fewer examples than $k$, and without loss of generality, assume A does not have example $e^l_k$.
       Namely, the positive and negative examples $e_1,...,e_t$ and $e'_1,...,e'_{t'}$ do not include $e^l_k$ (and any other example equivalent to $e^l_k$ with respect to $S$).
        We show that $e^l_k$ can be a positive example (i.e., $e^l_k\models \varphi_C$), without changing previous classifications of examples or literals and since $e^l_k \not \models R$, it implies that $l$ is not in $\varphi_C$ which contradicts A's classification of $l$.
        We set:
        $\varphi_C = \bigwedge_{Q\in (S \setminus S_N) \setminus S_{e^l_k}} Q$.
        $\varphi_C$ is consistent with the examples:
        \begin{itemize}[nosep,nolistsep]
          \item All previous positive examples satisfy $\varphi_C$: follows because every positive example satisfies all literals in $S \setminus S_N$ and $\varphi_C$ contains only predicates from $S \setminus S_N$.
         \item Every negative example $e'$ does not satisfy $\varphi_C$: 
           suppose $e' \models \varphi_C$ and suppose that there is another specification $\varphi_C'$ consistent with the classification of all the observed examples.
         For every two positive examples, literals that are satisfied by one but not by the other cannot be in $\varphi_C'$ (because otherwise one of them will not be a positive example). Thus, $e'$ cannot be a negative example because of one of the literals satisfied by one of the positive examples but not by another positive example, and there must be a literal in $\varphi_C'$ satisfied by all positive examples but not by $e'$, namely this literal is in $S\setminus S_N$.
         Consider all such literals in $S\setminus S_N$ that are not satisfied by $e'$. If all of them are in $S_{e_k^l}$, then $S_{e'}=S_{e_k^l}$, and $e' \equiv e_k^l$ in contradiction to our assumption that A has not observed such example. Thus, there is a predicate $Q\in S_{e'} \setminus S_{e_k^R}$ which is also in $\varphi_C'$.
         However in this case this predicate is also in $\varphi_C$, i.e., $e' \not \models \varphi_C$.

        \item ${e^l_k} \models \varphi_C$: since for every $Q \in \varphi_C$, $Q\notin S_{{e^l_k}}$ and thus ${e^l_k} \models Q$.
        \end{itemize}
        $\varphi_C$ is consistent with the literals:
        \begin{itemize}[nosep,nolistsep]
        \item For all $Q\in S_P$, $Q$ is in $\varphi_C$: follows since $S_P \subseteq S \setminus S_N$ and since $e^l_k \models S_P$ and thus $S_{e^l_k} \cap S_P =\emptyset$.
        \item For all $Q\in S_N$, $Q$ is not in $\varphi_C$: follows by construction.
        \end{itemize}
        However, $l$ is not in $\varphi_C$ (because it is not yet in $S_P$ and it belongs to $S_{e^l_k}$), and thus A classified $l$ incorrectly.
 \end{itemize}

 \para{\theref{predcomplexlinear} Proof.}
 Given this condition, every iteration of \algref{exam} is guaranteed to find a literal $l$ for which $sets$ consists of a single set, which implies that \algref{exam} returns one example to classify $l$, and this is true for every iteration of \alg. This follows since if the initial set in $sets$, which contains $\neg l$ (for C-\alg) or $l$ (for D-\alg), is satisfiable, then \algref{exam} completes (and returns the corresponding example), and otherwise if $sets$ is unsatisfiable, then this condition implies that all UNSAT cores contain exactly one predicate (excluding $l$ and the predicates from $S_P$) and thus $|sets|$ does not contain more than one set at any point.

 \para{Claim \ref{relativecomp} Proof.}
 First, note that at every iteration, $S_P$ is satisfiable (since all positive examples satisfy it). Also, every literal in $S\setminus S_N$ is not implied from $S_P$ because otherwise it would have been removed by the \scode{implied} operation. Thus, $\bigwedge_{Q \in S_P} Q \land \{\neg l\}$ is satisfiable. At each iteration of C-\alg, we can pick the literal $l$ in $S\setminus (S_N\cup S_P)$ pertaining to $x,y$ minimizing $|x-y|$ (if there are several, we pick any). The negation of $l$ may only require the negation of all literals over $x,z$ of values $z$ satisfying $S_P \models (y=z)$ and literals over $y,w$ of $w$ satisfying $S_P \models (x=w)$. For any other value $t$, if $t$ is not between $x,y$ then literals pertaining to $t$ are oblivious to negation of $l$, and if $t$ is between $x,y$ then predicates pertaining to $t$ and any value between $x,y$ (including $x,y$) were already classified, either to $S_P$ and thus can be classified along with $\neg l$, or to $S_N$ and thus need not be satisfied. Thus, overall the formulas in this claim are satisfiable.

  \subsection{Section 5}
  \para{\theref{dnfpred} Proof}
 We prove in induction that during the execution of Gen-\alg, $\bigvee_{l \in C_P} l \models \varphi_C \models \neg [\bigvee_{l \in C_N} l]$ and since at the end of the execution $=S_P$ and $\bigvee_{l \in C_P} l \equiv  \neg [\bigvee_{l \in C_N} l]$, it follows that $\bigvee_{l \in C_P} l \equiv \varphi_C$.
 
 Base: initially, $C_{P}=C_{N}=\emptyset$, and thus we have to show $false \models \varphi_C \models \neg[false]$, which clearly holds. 

Step:
We assume $\bigvee_{l \in C_P} l \models \varphi_C \models \neg [\bigvee_{l \in C_N} l]$ and show that updates to $C_{P}$ or $C_{N}$ preserve these implications. 
To this end, we rely on two claims:
\begin{inparaenum}[(i)]
  \item if given an example $e$ which belongs to exactly a single sub-concept, the modified C-\alg\ learns a conjunction which is not an over-generalization, and
  \item if given an example $e$ which belongs to more than one sub-concept, \scode{overgen} detects this, and thus Gen-\alg\ does not add its conjunction to $C_P$ or $C_N$.
\end{inparaenum}
Given the two claims, it follows that at each iteration, if the example submitted to C-\alg\ is positive, then its conjunction is satisfied only by positive examples, and thus extending $C_P$ preserves $\bigvee_{l \in C_P} l \models \varphi_C$, and similarly if the example is negative, then its conjunction is satisfied only by negative examples, and thus extending $C_N$ preserves $ \varphi_C \models \neg [\bigvee_{l \in C_N} l]$. 
We next prove the claims.

\begin{claim}
If an example $e$ belongs to exactly a single sub-concept, the modified C-\alg\ learns a conjunction which is not an over-generalization.
\end{claim}
Proof. 
To prove that its learned conjunction is not an over-generalization, we prove that it is not satisfied by any negative example (because a conjunction is an over-generalization only if it is satisfied by a negative example).
If $e$ belongs to exactly a single sub-concept, captured by a conjunction $c$, then it satisfies all the conjunction's constraints, and for each other sub-concept it has at least one constraint which $e$ does not satisfy.
Let $c'$ be the conjunction learned by C-\alg.
We prove that every example that satisfies $c'$ either satisfies $c$ or is a positive example satisfying a different sub-concept. Let $l$ be a literal in $c$. Modified C-\alg\ classifies $l$ either by generating examples for $l$ or if detecting that $l$ is implied by the learned conjunction. In the latter case, $c$ satisfies $l$, and thus $l$ is added to $S_P$ and the claim holds. We thus focus on the case where C-\alg\ generates examples to classify $l$. If $l$ can be classified using a single example $e_l$, then $e_l$ satisfies all constraints $e$ satisfies, except for $l$ whose negation is satisfied.
Since $l$ is part of $c$, $e_l$ does not satisfy this sub-concept. Also, since for any other sub-concept $e$ does not satisfy at least one constraint, $e_l$ also does not satisfy at least one constraint, too, and thus $e_l$ does not belong to any of the other sub-concepts. Thus, $e_l$ is classified as a negative example (or positive, in case C-\alg\ is given a negative example to generalize), and $l$ is classified to $S_P$. Otherwise, if $l$ cannot be classified with a single example, namely it has multiple examples $e_l^1,...,e_l^k$, each satisfies the negation of $l$. 
For every negative example (or positive, in case C-\alg\ is given a negative example to generalize), a disjunction eliminating it is added to $S_P$.
Note that the all these disjunctions contain $l$ and thus all positive examples of this sub-concept satisfy these disjunctions. Further note that adding the disjunctions enables at most the positive examples among $e_l^1,...,e_l^k$ to satisfy it. This follows since any other example negates an additional literal $l'$, which either cannot be satisfied with $c$ or that it is yet to be classified, and when it will be classified its negation will be tested and \alg\ will discover if negating it results in negative examples.



\begin{claim}
If an example $e$ belongs to more than one sub-concept then:
\begin{inparaenum}
  \item the modified C-\alg\ learns a conjunction which is an over-generalization, and \label{proofpointone}
  \item \scode{overgen} detects this.\label{proofpointtwo}
\end{inparaenum}
\end{claim}
Proof of \ref{proofpointone}:
If $e$ satisfies the constraints of (at least) two concepts, then when C-\alg\ negates the constraints of the first concept, the other concept's constraints are satisfied, resulting in observing only positive examples and thus adding these constraints to $S_N$. When C-\alg\ negates the constraints of the second concept, the constraints of the first concept are satisfied (because the second modification ensures that the literals in $S_N$ are satisfied), and thus the second concept's constraints also added to $S_N$, and thus $S_P$ is missing constraints of every sub-concept $e$ belongs to. This implies that the conjunction learned  must be an over-generalization. This is because if there are no negative examples satisfied by it, then there is no need in the sub-concept learned by C-\alg\ contains all the sub-concepts, in which case these sub-concepts are meaningless and thus can be ignored, and then $e$ would not have been considered as part of two sub-concepts.

Proof of \ref{proofpointtwo}:
Let $con$ be the conjunction learned by the modified C-\alg, namely $con$ over-generalizes $e$. Since there are (at least) two sub-concepts containing $e$ and captured by conjunctions that include $con$ and additional literals. These literals are:
  \begin{inparaenum}[(i)]
    \item satisfied by $e$, and
    \item not implied by $con$.
  \end{inparaenum}
  Also, since $con$ is an over-generalization, there are negative examples satisfying it. Namely, there are examples satisfying $con$ but not any of the conjunctions of the sub-concepts. In particular, there are negative examples:
  \begin{inparaenum}[(i)]
    \item satisfying $con$, and
    \item not satisfying some of the other literals satisfied by $e$ (which are not implied by $con$),
  \end{inparaenum}
  In addition, any example satisfying even fewer constraints satisfied by $e$, is also a negative example.
  Thus, since \scode{overgen} constructs all the examples not satisfying a maximal number of constraints, it must encounter a negative example and determine that $con$ is an over-generalization. 
  
  \para{\theref{predcomplex2} Proof} 
  Let A be a DNF-\name\ algorithm (i.e., an algorithm that can learn arbitrary DNF formulas), $D$ a domain, $S$ predicates over $D$ of size $n$, an unknown target formula $\varphi_C$, $E_P$ and $E_N$ sets of positive and negative examples.
  We first prove that the modified C-\alg\ generalizes examples as much as possible:
  \begin{claim}
    Let $c$ be a conjunction in $\varphi_C$, C-\alg\ captured $c$ via $c_{\alg}$ which is satisfied by at least all the examples satisfied by $c$.
    \end{claim}
    Proof of claim. Assume that C-\alg\ generalizes from an example $e$ and assume in contradiction that there is an example $e'$ satisfying $c$ but not $c_{\alg}$. This means that there is a literal $l$ such that $e' \not \models l$ for which either $l'$ is in $c_{\alg}$ or in a disjunction in $c_{\alg}$. In the first case, it means that C-\alg\ has added $l$ after generating an example $e_l$ satisfying all literals from $S$ that $e$ satisfies (except for $l$) and $\neg l$, and this example was classified as negative by the user (or positive if we generalize from a negative example). Since $c$'s literals must be a subset of the literals from $S$ satisfied by $e$, it follows that $e_l$, which is a negative example, satisfies $c$, but this cannot happen since $c$ is a correct sub-concept, i.e., includes only positive examples. In the latter case, $c_{\alg}$ excluded exactly the negative behaviours, and thus if $e'$ is excluded, then the negative example $e_N$ that led to adding the disjunction that excludes $e'$, implies that $e'$ cannot be in $c$: if it were, then since it satisfies fewer literals than $e_N$ (compared to the original example $e$), $e_N$ also must be in $c$, but it was classified as negative, in contradiction.
    
    The next claim states that modified C-\alg\ presents a minimal number of questions, given the assumption that positive examples may be examples which are not part of the hidden concept.
    \begin{claim}
    If positive classifications of examples do not imply that the examples are part of the learned sub-concept, 
      then the modified C-\alg\ generates a minimal number of questions to learn the sub-concept.
    \end{claim}
    Proof of claim. C-\alg\ was shown to present a minimal number of questions. Compared to it, the modified C-\alg\ introduces more questions only when there is dependency because it does not ignore literals in $S_N$ when generating the formulas $\psi$ (in \scode{getMinLiteralNExamples}). To prove the claim, we demonstrate that ignoring these literals may result in incorrect specifications, and thus the additional questions cannot be avoided. Consider the domain of boolean vectors of size $3$, the predicates are monomials, i.e., $x_0,x_1,x_2$, and the specification is: $(x_0 \land x_1)\vee(x_0 \land x_2)$. Suppose that C-\alg\ is given the example $(1,1,0)$ which satisfies the first conjunction but not the second, and thus C-\alg\ should be able to generalize it correctly (\scode{overgen} does not detect such over-generalizations). Initially, C-\alg\ tests whether $\neg x_2$ is part of the conjunction, and presents the example $(1,1,1)$, which is positive, and thus it infers that $\neg x_2$ is not in the conjunction. Next, it tests whether $x_1$ is in the specification. Since $\neg x_2$ is ignored, an example satisfying $x_0$ and $\neg x_1$ is $(1,0,1)$ which is positive (because it satisfies the other conjunction), and thus $x_1$ erroneously is removed from the conjunction.
  
    The next claim states that any algorithm that does not generalize both positive and negative examples may present is outperformed by Gen-\alg.
    \begin{claim}
      Let A be a DNF-\name. If A only learns a single DNF formula (satisfied by the positive examples only, or by the negative examples only), then A may present $\Omega(2^n)$ questions, where $n$ is the size of $S$ (the set of literals).
    \end{claim}
    Proof. Suppose A only learns a DNF formula satisfied by the positive examples and consider a concept containing exactly a single example from the domain. Then, A has to examine all examples in the domain, or more precisely all non-equivalent examples with respect to the predicates in $S$. In general, there are $\Omega(2^n)$ such examples. Even if there are some dependencies and not all combinations of literals from $S$ are satisfiable, in general this number is still exponential.
    
    In contrast, Gen-\alg\ will generalize the negative examples with a minimal number of questions, and in particular will not present more questions than A.
    
The next claim states that if A generalizes a conjunction, that is by dropping its constraints, then it risks in over-generalization and thus has to trigger at least the questions that \scode{overgen} introduces to discover this.
\begin{claim}
  If at some point of the algorithm A adds to the learned DNF the conjunction $con = \bigwedge_{l\in S'}l $ where $S'\subseteq S$, it must observed all examples \scode{overgen} would generate for $con$ with the positive example $e$ used by C-\alg\ to compute $con$.
\end{claim}
Proof. If A generated this conjunction and added it to a DNF, it must have seen a positive example $e$ satisfying all $con$'s literals (otherwise the DNF specification is incorrect). If A does not examine one of the examples \scode{overgen} generates for $e$ and $con$, then this example may be negative. 
This follows because the examples \scode{overgen} generates do not imply one another nor are implied by other examples satisfying $con$, and thus A could not avoid this example through a different example.
Namely, A added to the DNF a cube that is satisfied by a negative examples, and thus learned an incorrect DNF.

  Theorem Proof. The three first claims imply that the questions submitted to C-\alg\ are of a minimal number, the examples C-\alg\ presents are of a minimal number, and the learned conjunctions are guaranteed to cover as many examples from the domain as possible.
The last claim implies that A cannot reduce the number of questions \scode{overgen} presents. 
  Thus, A may only have an advantage over Gen-\alg\ if it happened to pick better examples to generalize. 
  However, A (as Gen-\alg) has no knowledge on the unlearned sub-concepts, and if for some concept it ``was lucky'' to pick an example $e$ not part of two sub-concepts, and Gen-\alg\ picked an example $e'$ that belongs to more than one sub-concept, then there are concepts which have the same sub-concepts as A and Gen-\alg\ learned so far, in which the example $e$ is part of two sub-concept and $e'$ is not. Thus, overall Gen-\alg\ learns concepts with a minimal number of questions.
  
  
\section{Technical Analysis Common Patterns}\seclabel{patterns}
In this section we describe the patterns used to evaluate C-\alg\ in~\secref{techanalysis}. We used the following patterns:
\begin{inparaenum}[(i)]
  \item head and shoulders,
  \item cup with handle,
  \item double tops,
  \item symmetrical triangle,
  \item rectangle, and
  \item flag.
\end{inparaenum}
The last five patterns are illustrated in \figref{patterns}; for further reading see ~\cite{bulkowski2012visual,EncPatterns,Investopedia}.

The challenge in evaluating C-\alg\ is to decide on the pattern definition to use as pattern definitions are subjective.
 To overcome this challenge,
we ran several experiments for each pattern, each with a different formula (but with the same example). The different definitions, taken from textbooks and online forums, span a range of possible definitions, from the most permissive to the most restrictive.
We next provide a general description of the patterns and the definitions used.

\begin{figure}[t]
\centering
\begin{tabular}{p{1.4cm}p{2.8cm}p{2.8cm}}
{\bf \hspace{0.1cm} Pattern} & {\bf \hspace{1cm} Figure} & {\bf \hspace{0.2cm} Example Chart}\\
Cup with Handle &
\raisebox{-1.45cm}{\includegraphics[height=2cm, width=3cm,clip = true, trim = 0pt 35pt 20pt 22pt]{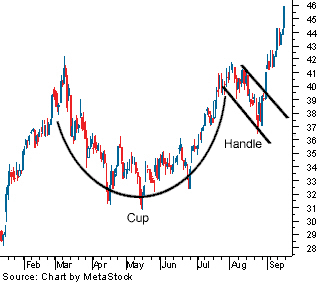}} &
\raisebox{-1.45cm}{\includegraphics[height=2cm, width=3cm,clip = true, trim = 0pt 0pt 0pt 0pt]{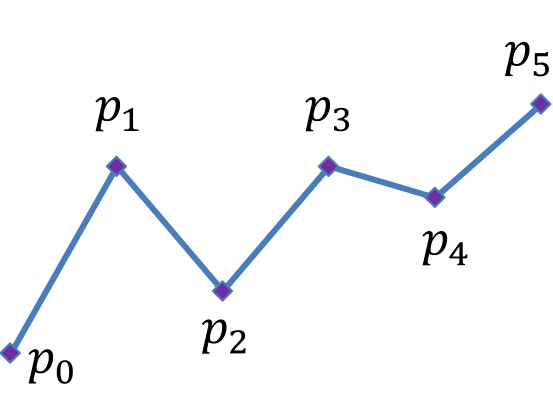}} \\
Two Tops &
\raisebox{-1.45cm}{\includegraphics[height=2cm, width=3cm,clip = true, trim = 0pt 0pt 0pt 22pt]{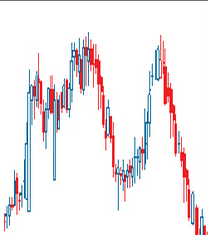}} &
\raisebox{-1.45cm}{\includegraphics[height=2cm, width=3cm,clip = true, trim = 0pt 0pt 0pt 0pt]{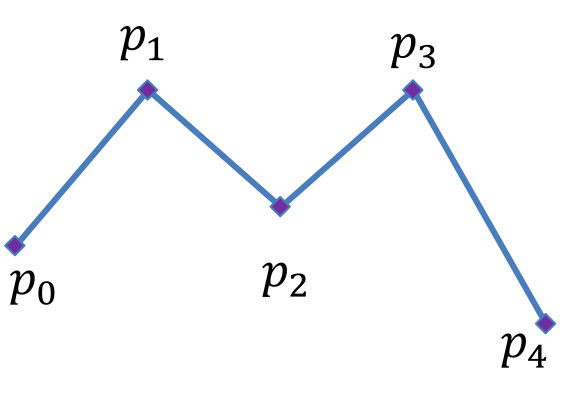}}\\
Symmetrical Triangle&
\raisebox{-1.8cm}{\includegraphics[height=2cm, width=3cm,clip = true, trim = 4pt 14pt 13pt 35pt]{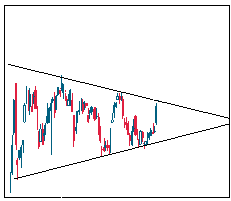}} &
\raisebox{-1.45cm}{\includegraphics[height=2cm, width=3cm,clip = true, trim = 0pt 0pt 0pt 0pt]{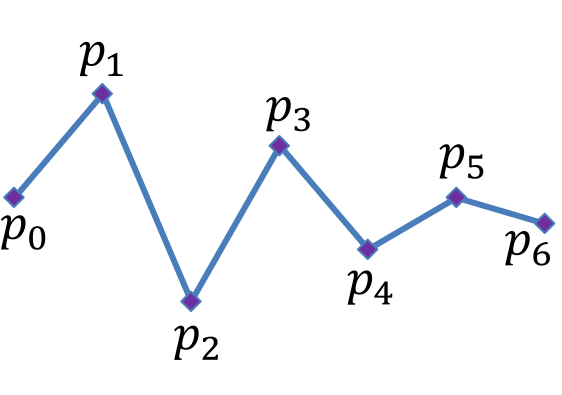}}\\
Flag &
\raisebox{-1.8cm}{\includegraphics[height=2cm, width=3cm,clip = true, trim = 30pt 40pt 10pt 15pt]{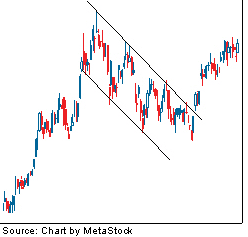}} &
\raisebox{-1.45cm}{\includegraphics[height=2cm, width=3cm,clip = true, trim = 0pt 0pt 0pt 0pt]{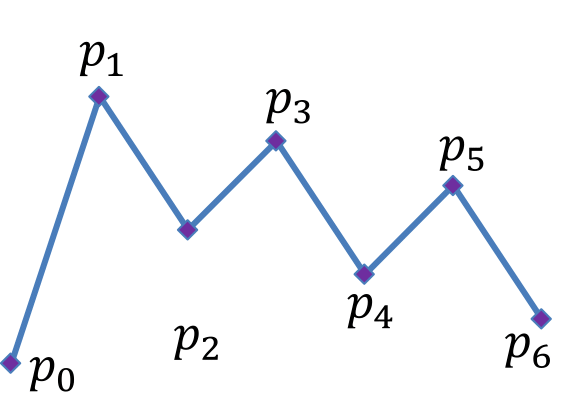}}\\
Rectangle &
\raisebox{-1.45cm}{\includegraphics[height=2cm, width=3cm,clip = true, trim = 0pt 30pt 18pt 15pt]{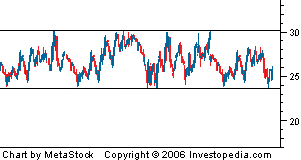}} &
\raisebox{-1.45cm}{\includegraphics[height=2cm, width=3cm,clip = true, trim = 0pt 0pt 0pt 0pt]{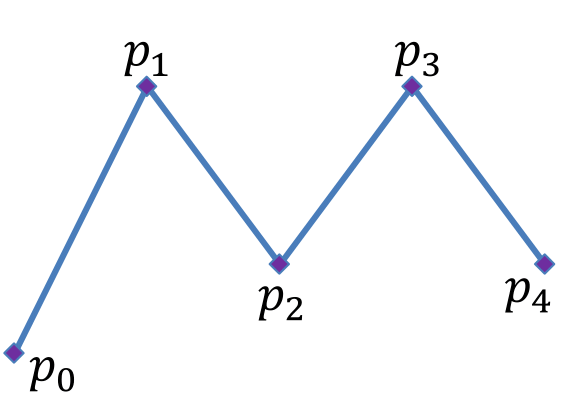}}\\

\end{tabular}
\caption{The New Patterns (figures taken from~\cite*{Investopedia}).}\figlabel{patterns}
\end{figure}

\vspace{3pt}\noindent{\textit{Head and Shoulders}} Three peaks, the middle is the highest.
\begin{enumerate}[(1),nolistsep]
  \item Most permissive -- three peaks,  middle one is the highest.\label{head2}
  \item \ref{head2} with shoulders higher than all lows. 
      \label{head3}
  \item \ref{head3} where $p_0,p_6$ are lower than the other points.\label{head4}
  \item \ref{head4} with ascending ``neckline'' ($p_0$$\prec$$p_2$$ \prec$$p_4$) and $p_6$$\prec$$ p_0$.\label{head5}
    \item Most restrictive -- the given chart is the only valid chart. \label{head1}
\end{enumerate}

\vspace{3pt}\noindent{\textit{Cup with Handle}} A rise, followed by a cup-shape, then a decline (``the handle''), and finally another rise.
 \begin{enumerate}[(1),nolistsep]
  \item Most permissive -- all four parts exist.\label{cup1}
  \item \ref{cup1} with significant rise: $p_5$ is higher than the other points.\label{cup2}
  \item \ref{cup2} with $p_0$ lower than the other points.\label{cup3}
  \item \ref{cup3} with handle not lower than the cup ($\neg(\ltpred{p_4}{p_2})$).\label{cup4}
    \item Most restrictive -- the given chart is the only valid chart. \label{cup5}
\end{enumerate}
\vspace{3pt}\noindent{\textit{Two Tops}} Two peaks of equal height.
 \begin{enumerate}[(1),nolistsep]
  \item Most permissive -- there are two equal height tops.\label{top1}
  \item \ref{top1} with middle low ($p_2$) not lower than the other lows.\label{top2}
  \item \ref{top2} with last point ($p_4$) lower than the other points.\label{top3}
    \item Most restrictive -- the given chart is the only valid chart. \label{top4}
\end{enumerate}
The next patterns are captured by constraints that leave little room for different definitions and thus only two are listed.

\vspace{3pt}\noindent{\textit{Symmetrical Triangle}} Descending peaks ($p_1$$\succ$$p_3$$\succ$$p_5$), ascending lows ($p_2$$\prec$$p_4\prec p_6$), and $p_2$$\prec$$p_0$,
$p_0$$\prec$$p_1$.
 \begin{enumerate}[(1),nolistsep]
  \item Most permissive -- $p_0$ appears between $p_1$ and $p_2$.\label{trian1}
    \item Most restrictive -- the given chart is the only valid chart. \label{trian2}
\end{enumerate}
\vspace{3pt}\noindent{\textit{Flag}} A pole followed by descending peaks ($p_1$$\succ$$p_3$$\succ$$p_5$), descending lows ($p_2$$\succ$$p_4$$\succ$$p_6$), and $p_0$ lower than all points.
 \begin{enumerate}[(1),nolistsep]
  \item Most permissive -- $p_2$ and $p_5$ may be equal.\label{flag1}
    \item Most restrictive -- the given chart is the only valid chart. \label{flag2}
\end{enumerate}
\vspace{3pt}\noindent{\textit{Rectangle}} Peaks ($p_1,p_3$) are equal, lows ($p_2,p_4$) are equal, and $p_0$ not higher than $p_1$.
 \begin{enumerate}[(1),nolistsep]
  \item Most permissive -- $p_0$ is not higher than $p_1$.\label{rec1}
    \item Most restrictive -- the given chart is the only valid chart. \label{rec2}
\end{enumerate}

\section{Full results for C-SPEX}\applabel{extendedresults}
In this section, we provide the full results of~\secref{techanalysis} that include the overall time (in milliseconds).
\begin{table*}
  \centering
\footnotesize
    \begin{tabular}{rrrrrrccrr}
    \toprule
          &       &       &       & \multicolumn{4}{c}{\textbf{\#Questions}} & \multicolumn{2}{c}{\textbf{Time (millisec)}} \\
    \midrule
          &       &       &       & \textbf{CSPEX} & \multicolumn{3}{c}{\textbf{Oracle Based}} & \multicolumn{1}{c}{\textbf{CSPEX}} & \multicolumn{1}{c}{\textbf{OB}} \\
    \textbf{Pattern } & \boldmath{}\textbf{ $|S_{0}|$ }\unboldmath{} & \textbf{ Def. } & \boldmath{}\textbf{ $|S_{P}|$ }\unboldmath{} & \multicolumn{1}{c}{\textbf{Num}} & \textbf{ Avg. } & \textbf{ Max } & \textbf{ Min} &       &  \\
          & \multicolumn{1}{c}{\multirow{5}[2]{*}{42}} &  \ref{head2}  & 6     & \multicolumn{1}{c}{18} & 44.4  & 57    & 38    & 360   & 139 \\
    Head and   & \multicolumn{1}{c}{} &  \ref{head3}  & 10    & \multicolumn{1}{c}{18} & 54.8  & 61    & 43    & 223   & 142 \\
    Shoulders & \multicolumn{1}{c}{} &  \ref{head4} & 10    & \multicolumn{1}{c}{17} & 61.1  & 88    & 43    & 344   & 155 \\
          & \multicolumn{1}{c}{} &  \ref{head5} & 7     & \multicolumn{1}{c}{14} & 50.6  & 71    & 40    & 285   & 132 \\
          & \multicolumn{1}{c}{} &  \ref{head1} & 6     & \multicolumn{1}{c}{12} & 58.4  & 77    & 37    & 267   & 147 \\
          & \multicolumn{1}{c}{\multirow{5}[2]{*}{30}} &  \ref{cup1} & 5     & \multicolumn{1}{c}{12} & 38.3  & 48    & 27    & 107   & 82 \\
    Cup with   & \multicolumn{1}{c}{} & \ref{cup2} & 6     & \multicolumn{1}{c}{12} & 43.6  & 57    & 33    & 93    & 82 \\
    Handle & \multicolumn{1}{c}{} &  \ref{cup3} & 7     & \multicolumn{1}{c}{13} & 35.1  & 41    & 28    & 232   & 70 \\
          & \multicolumn{1}{c}{} &  \ref{cup4} & 7     & \multicolumn{1}{c}{13} & 36.4  & 41    & 32    & 184   & 66 \\
          & \multicolumn{1}{c}{} &  \ref{cup5} & 5     & \multicolumn{1}{c}{10} & 38.6  & 45    & 22    & 111   & 69 \\
    Two Tops  & \multicolumn{1}{c}{\multirow{4}[2]{*}{20}} &  \ref{top1} & 6     & \multicolumn{1}{c}{9} & 19.4  & 20    & 18    & 56    & 29 \\
          & \multicolumn{1}{c}{} & \ref{top2} & 6     & \multicolumn{1}{c}{9} & 18.3  & 20    & 16    & 49    & 29 \\
          & \multicolumn{1}{c}{} &  \ref{top3} & 6     & \multicolumn{1}{c}{7} & 18.7  & 19    & 17    & 56    & 28 \\
          & \multicolumn{1}{c}{} &  \ref{top4} & 4     & \multicolumn{1}{c}{6} & 17    & 17    & 17    & 56    & 25 \\
    Symmetrical    & \multicolumn{1}{c}{\multirow{2}[2]{*}{42}} &  \ref{trian1} & 7     & \multicolumn{1}{c}{16} & 71.4  & 76    & 62    & 23    & 186 \\
    Triangle & \multicolumn{1}{c}{} &  \ref{trian2} & 7     & \multicolumn{1}{c}{14} & 66.5  & 78    & 37    & 239   & 165 \\
    Flag  & \multicolumn{1}{c}{\multirow{2}[2]{*}{42}} & \ref{flag1} & 7     & \multicolumn{1}{c}{17} & 64    & 89    & 52    & 706   & 175 \\
          & \multicolumn{1}{c}{} &  \ref{flag2} & 6     & \multicolumn{1}{c}{16} & 54.3  & 90    & 43    & 67    & 136 \\
    Rectangle    & \multicolumn{1}{c}{\multirow{2}[2]{*}{20}} &  \ref{rec1} & 6     & \multicolumn{1}{c}{8} & 27.4  & 28    & 25    & 78    & 36 \\
          & \multicolumn{1}{c}{} &  \ref{rec2} & 6     & \multicolumn{1}{c}{8} & 25.2  & 27    & 22    & 72    & 47 \\
    \bottomrule
    \end{tabular}%
  \caption{Results for CSPEX. Number of questions presented by C-\alg\ vs. average, maximal, and minimal number of questions presented by the Oracle-Guided approach. Time for generating the next question for both algorithms is shown in milliseconds.}. \tablabel{appendix-cspex}%
\end{table*}%

\end{document}